\documentclass[twocolumn,linenumbers,nopacs]{revtex4}

\usepackage{graphicx}%
\usepackage{dcolumn}
\usepackage{amsmath}
\usepackage{float}

\makeatletter
\def\btt#1{\texttt{\@backslashchar#1}}%
\DeclareRobustCommand\bblash{\btt{\@backslashchar}}%
\makeatother

\begin{document}

\title{Quantum correlation interferometry in reflection: application to the quantum-classical transition, decoherence, and indirect measurement}

\author{F.V. Kowalski}
\affiliation{Physics Department, Colorado
School of Mines, Golden CO. 80401 U.S.A.}

\begin{abstract}
A many-body interferometer is described in which all of its components are treated as quantum objects. It consists of particles reflecting elastically from  a ``mirror.'' Quantum correlation is a consequence of conservation of energy and momentum while interference occurs when the order in which the non-local particles reflect is indeterminate. The resulting superposition exhibits correlated interference with diverse characteristics depending on the structure of the many-body wavegroup. Two non-local microscopic particles reflecting from a mesoscopic ``mirror'' illustrate unique features of this correlation interferometer. The microscopic momentum exchanged then results in small displacements of the superposed mesoscopic mirror substates, which mitigates experimental difficulties in determining the quantum-classical boundary. Quantum behavior of this mesoscopic mirror, evident in indirect measurements involving correlations between only the reflecting microscopic particles, disappears for a classical mirror which cannot exist in such superposition states.  
\end{abstract}


\maketitle

\section{Introduction}
\label{sec:intro}

Observation of macroscopic quantum phenomena continues to be a topic of interest \cite{schlosshauer0,arndt}. The systems which are studied can be placed into a simple although imperfect dichotomy. The first category involves the collective behavior of large numbers of interacting particles, such as that found in superconductors, superfluids, Bose-Einstein condensates, the quantum Hall effect, and proton tunneling in ice \cite{chen}. The second, which is the focus of the discussion below, involves quantum effects associated with center of mass motion of an object composed of a large number of atoms. This number covers three size scales: large molecules traversing an interferometer \cite{Hackermuller}, mesoscopic mechanical oscillators \cite{ock}, and macroscopic masses (e.g. used in gravity wave detectors) \cite{chen1,danilishin}.

An implicit goal is to address the measurement problem \cite{measurement}. The challenge is in reducing environmental coupling to the system being studied since such coupling can lead to measurement like interactions which decohere the quantum state. Therefore, observing such decoherence is sometimes an explicit research goal \cite{bose,penrose}. 

Predictions of such quantum phenomena vary in complexity. The simplest model is of one-body quantum interference where the interferometer traversed is described as a classical object. The next level of complexity involves many-body correlations, a simple two-body description of which is given by Bohm's version of the EPR paradox \cite{bohm}. 

Quantum correlated interference is a combination of these and can be divided into three categories. The first, type A or post-correlation interferometry, is of distinguishable particles which are correlated before they traverse interferometers. The beamsplitters and mirrors then act as potentials to split or guide the particles through the interferometers. These are assumed to be in the classical domain and are therefore referred to as classical potentials \cite{classical}.  Only the incident correlated particles are measured beyond the interferometer output ports. A simple example involves two microscopic particles which are generated in a decay and therefore are correlated via conservation of energy and momentum \cite{Gottfried}. They then each traverse a different double slit interferometer, resulting in correlated fringe patterns.

The second, type B, involves both initially uncorrelated particles and interferometers which become correlated when the particles traverse the interferometers. The mirrors, beamsplitters, and even the particles then become integral parts of the interferometers and therefore are all treated as quantum objects.  A simple example is of one non-local particle reflecting from a static mirror, both of which are initially uncorrelated, generating two-body correlated ``standing wave'' interference between both the particle and mirror \cite{kowalski1}. This interference is a consequence of the system being in a superposition of both having and not having reflected the particle. It is difficult to distinguish one or the other body as the mirror and therefore identify one as the interferometer and the other as the object which exhibits interference after traversing the interferometer. 

Rather than verifying quantum correlated interference between the particle and interferometer in type B systems, only the particle traversing the interferometer is typically measured at its output port. How such correlation is masked in these measurements is a topic discussed below. If the interferometer components behave as classical objects, which cannot exist in superposition states, then correlation between the interferometer and particle traversing it does not exist in principle. Nevertheless, this particle remains in a a superposition state associated with the different paths through the interferometer and therefore only it experiences interference. It is also possible in type B systems for some of the interferometer components to behave as quantum objects while others are in the classical domain generating correlation only between the particle and these quantum components.

Since quantum correlations are typically difficult to measure, attempts are made to find many-body quantum effects which are manifest in measurements on only a subset of the bodies, predicted by a marginal probability density function, PDF, (or a reduced density operator determined from a partial trace over the density matrix). This provides an indirect measurement of the quantum system. However, the full quantum correlation cannot be verified explicitly in a marginal PDF.

Fewer type A systems are used since it is difficult to generate correlated particles, inject them into an interferometer, and then measure their correlation at the interferometer output ports. Experiments which support type A models are a consequence of either the interferometers being in the classical domain or of marginal PDFs (measuring only the correlated particles at the output port) masking a correlation with the interferometer, an example of which is discussed in sec. \ref{sec:compareInt}. The latter is an example of type C systems. These involve initially correlated particles which experience additional correlation with the interferometer as they traverse it, an illustration of which is mentioned in sec. \ref{sec:expt}. 

The central example of correlated interference presented here, type B, involves two non-local particles interacting elastically only with a ``mirror,'' all three of which are treated as quantum objects and all are initially uncorrelated. The particles and mirror move in one dimension. Interference is a consequence of the reflection order being indeterminate. This is an example of an open interferometer for which the correlation is particularly difficult to verify experimentally. To mitigate this, such systems are transformed to closed interferometers facilitating indirect interferometric measurements via probes (the reflecting particles) whose behavior is predicted by a marginal PDF. A description of open and closed many-body interferometers is given in sec. \ref{sec:closedopen} while methods to transform open to closed interferometers are discussed in sec. \ref{sec:coupled}.

The calculations presented here differ from previous work in the following ways. First, all of the interferometer components are treated as quantum objects. Second, microscopic bodies reflecting from a mesoscopic mirror are used as an example of the results. It is then shown that decoherence can be mitigated in this system. Third, the importance of the many-body wavegroup structure on correlated interference is emphasized in these ab initio calculations. Fourth an indirect measurement is outlined in which the interference of one particle is correlated with that of the other via the quantum behavior of the unmeasured mesoscopic mirror, with which both microscopic particles interact. The results from this example are then generalized to predict indirect measurements in other coupled many-body interferometers, with a particular focus on microscopic particles reflecting from a mesoscopic mirror in closed interferometers. 

If the mirror behaves as a classical object it can no longer be in the superposition state associated with indeterminate reflection sequences of the particles. The quantum-classical boundary is then reached when correlated interference, which depends on this superposition state, disappears. This boundary has been a topic of investigation since the founding of quantum theory, which is indicative of the experimental difficulties involved in its resolution. Mitigation of some of these issues using many-body correlation interferometry in reflection is addressed below.

\section{Overview}
\label{sec:overview}

Before describing the calculations in detail, the following subsections give heuristic descriptions of (1) correlation and interference, (2) the many-body treatment of interference in reflection, (3) how mirror fringes of mesoscopic extent are maintained even as the mirror mass increases, (4) how the interaction of the environment with the mesoscopic mirror, which is in a superposition state, is mitigated, (5) why the interferometric correlations exist (described in sec. \ref{sec:closedopen}), (6) wavefunction collapse in a many-body interferometer, (7) the effect of wavegroups on correlated interference, (8) closed and open interferometers, (9) experimental evidence for type B correlation interferometry generated via recoil, and (10) the difference between simultaneous and asynchronous measurements in correlated interference. An outline is then given of the calculations that follow.

\subsection{Correlation and interference}
\label{sec:entanglement}

The PDF for a type A interferometer depends on, among other things, the number of particles that traverse the interferometer and the number of paths within it \cite{greenberger}. For bipartite systems, photon pairs are correlated by parametric downconversion before they traverse different interferometers for each photon \cite{greenberger,franson,pan,pan2}. However, examples of more than two-body type A systems are scarce, typically only involving gedankenexperiments \cite{greenberger,Gottfried}. There appears to be little discussion of type C systems.

Here the splitting mechanism involves elastic retro-reflection which is described by the motions of the initially uncorrelated particles and mirror centers of mass in one dimension (quantum scattering in molecular two-body systems has been well studied \cite{child}). This type B interference is a consequence of reflection both having or not having occurred, the uncertainty of which is due to the non-local character of one or more of the reflecting bodies. Correlation is then a consequence of conservation of energy and momentum in reflection. These fundamental principles, applied to recoil in other examples, have played an important role in attempts to explain Bohr's principle of complementarity \cite{wiseman}, although not without controversy \cite{drezet}.

\subsection{Interferometry in reflection}
\label{sec:reflection}

Interferometry in reflection \cite{Wiener} differs from division of wavefront and division of amplitude interferometers. One advantage, useful in studying the quantum-classical boundary, is that a mesoscopic mirror neither has to fit through slits nor traverse beamsplitters to exhibit interference. Another is that the difficulty in interferometer alignment is mitigated.

Yet another advantage is that reflection of a non-local microscopic particle, of mass $m$, can result in mirror fringe spacings which do not decrease with increasing mesoscopic mirror mass, $M$. This is a consequence of the superposition of mirror substates whose difference in momenta, $\hbar \Delta K$, is microscopic due to having and not having reflected the non-local microscopic particle. The resulting two mirror substate wavefunctions, of different wavevectors, then interfere with a phase difference $\Delta \phi_{refl}=\Delta K x \approx 2mvx/\hbar$, where $v$ is the speed of microscopic mass $m$, and therefore a fringe spacing $2 \pi/\Delta K$, which is independent of $M$. 

These mirror fringe spacings are to be contrasted with those from a quantum object of mass M after it has traversed a ``classical" double slit potential, where the superposition is between waves which have traveled different distances to the observation point, resulting in a phase difference given by $\Delta \phi_{slits}= K \Delta x$, where in the far field limit $\Delta x=d \sin \theta$, with $d$ being the slit spacing and $\theta$ the diffraction angle. The fringe spacing, in this one-body treatment of the body as it traverses a double slit, is then inversely proportional to M, resulting in fringe spacings so small that they are imperceptible for large $M$. This has similar implications for Kapitza-Dirac-Talbot-Lau interferometers, which have been used to demonstrate quantum interference with large molecules \cite{gerlich}.

Such a one-body treatment of double slit interference has an analogy with one-body interference in reflection which involves a particle reflecting from a classical fixed potential ``mirror." In this case, the standing wave fringes of the particle also decrease in size with increasing particle mass. 

However, in a two-body quantum treatment of this particle-mirror system, fringes are also introduced in the mirror substate. Now, rather than increasing the mass of the reflecting particle (or its momentum) let it remain microscopic while only the mass of the mirror increases. Reflection of a microscopic particle from a mesoscopic mirror in this two-body treatment then reveals quantum effects in the mesoscopic mirror that are normally associated with the microscopic domain via the superposition of mirror substates of different {\em microscopic} momenta. That is, the mirror fringes do not become imperceptibly small as the mirror mass increases. Correlation, a consequence of conservation of energy and momentum, then influences the fringe locations in the two-body PDF as a function of the particle and mirror coordinates (as shown in fig. \ref{fig:2body}). 

Interferometry with massive particles is often associated with an enhanced sensitivity to path differences (e.g. $\Delta x$) compared with that of optical methods, as is illustrated by the decreasing fringe spacing in the double slit example given above for increasing $M$. That is, $\partial \phi_{slits}/\partial \Delta x=M V/\hbar$, where $V$ is the speed of mass $M$, vs. $\partial \phi_{optical}/\partial \Delta x=k$ for an optical system, where $k$ is the photon wavevector. 

Constructing interferometers which are insensitive to vibrations and alignment becomes more difficult with this increased sensitivity. However, for interference of the mesoscopic mirror of mass M in two-body reflection of a microscopic particle $\partial \phi_{refl}/\partial x= 2mv/\hbar$. Practical methods to determine just that a mesoscopic body is in a superposition state, rather than using it to interferometrically measure small path displacements, require reduced rather than enhanced such sensitivity. This can be implemented in two-body reflection interferometry for a microscopic body reflecting from a mesoscopic mass since then $\partial \phi_{refl}/\partial x <<\partial \phi_{slits}/\partial \Delta x$. 

Measurements of non-zero rest mass particle reflection have involved mirrors that reflect atoms \cite{kouznetsov} and Bose-Einstein condensates \cite{pasquini}, atoms reflecting from a solid surface \cite{shimizu}, neutrons \cite{hils} and atoms \cite{colombe} reflecting from vibrating mirrors, and atoms reflecting from a switchable mirror \cite{szriftgiser}. While some of these results involve interferometry, none focus on correlation interferometry in reflection. Nevertheless, they provide evidence that coherence can be maintained in reflection of massive particles from mesoscopic to macroscopic masses. Reflection of photons, discussed in sec. \ref{sec:discussion}, is ubiquitous. 

\subsection{Interaction with the environment}
\label{sec:environment}

In addition to generating mesoscopic mirror fringe spacings, another advantage of a microscopic particle of mass $m$ reflecting from a mesoscopic mirror of mass $M$ is that the centers of mass of these interfering mirror substates (having and not having reflected the particle) are displaced by the small distance, $\Delta D_{2body} \approx 2 m (v-V) t/M$ where $v$ and $V$ are the particle and mirror velocities respectively. This displacement of the superposed states is an important parameter in models of quantum decoherence, which fall into two limiting cases  \cite{schlosshauer}:
(a) the decohering environmental particle has a wavelength $\lambda_{E}<<\Delta D_{2body}$ thereby allowing path information to be encoded in each reflection of the environmental particle or (b) $\lambda_{E}>>\Delta D_{2body}$ which requires many interactions with environmental particles to determine path information. 

The wavelength of a typical environmental particle is much larger than $\Delta D_{2body}$ for a microscopic particle reflecting from a mesoscopic mirror and therefore model (a) is not applicable \cite{kowalski1}. The more likely decoherence mechanism (b) is characterized by an exponential decay of the off-diagonal density matrix element terms with a time constant which depends on $(\Delta D/\lambda_{T})^{2}$, where $\lambda_{T}$ is the thermal deBroglie wavelength of the mirror \cite{zurek}. In this model, the quick quantum decoherence of the Schr\"odinger cat states, for the superposed cats being separated by a few centimeters, is a consequence of this separation being much greater than the cat thermal wavelength. However, $\Delta D_{2body}$ is much less than $\lambda_{T}$ for a mirror, of a mass similar to that of the cat, reflecting a microscopic particle. Therefore this mechanism of environmental decoherence can in principle be mitigated with correlated interference in reflection and so holds potential for studying the quantum-classical boundary. This issue is revisited in section \ref{sec:discussion}. 

Decoherence due to thermal radiation emitted by fullerene molecules traversing a Talbot-Lau interferometer has been observed \cite{Hackermuller}. The wavelength of the thermal radiation emitted by the molecule could be made less than the mesoscopic separation of the molecular center of mass superposed states (of order $1 \mu m$). Similarly, decoherence of fullerene molecules, due to gas atoms collisions has been measured \cite{hornberger}. Again, the center of mass displacement of the fullerene superposed states was mesoscopic. These results provide experimental evidence for the models of decoherence described above, but only for microscopic masses.

\subsection{Three-body interferometer}
\label{sec:interferometer}

\begin{figure} [H]
\centering
\includegraphics[scale=0.3]{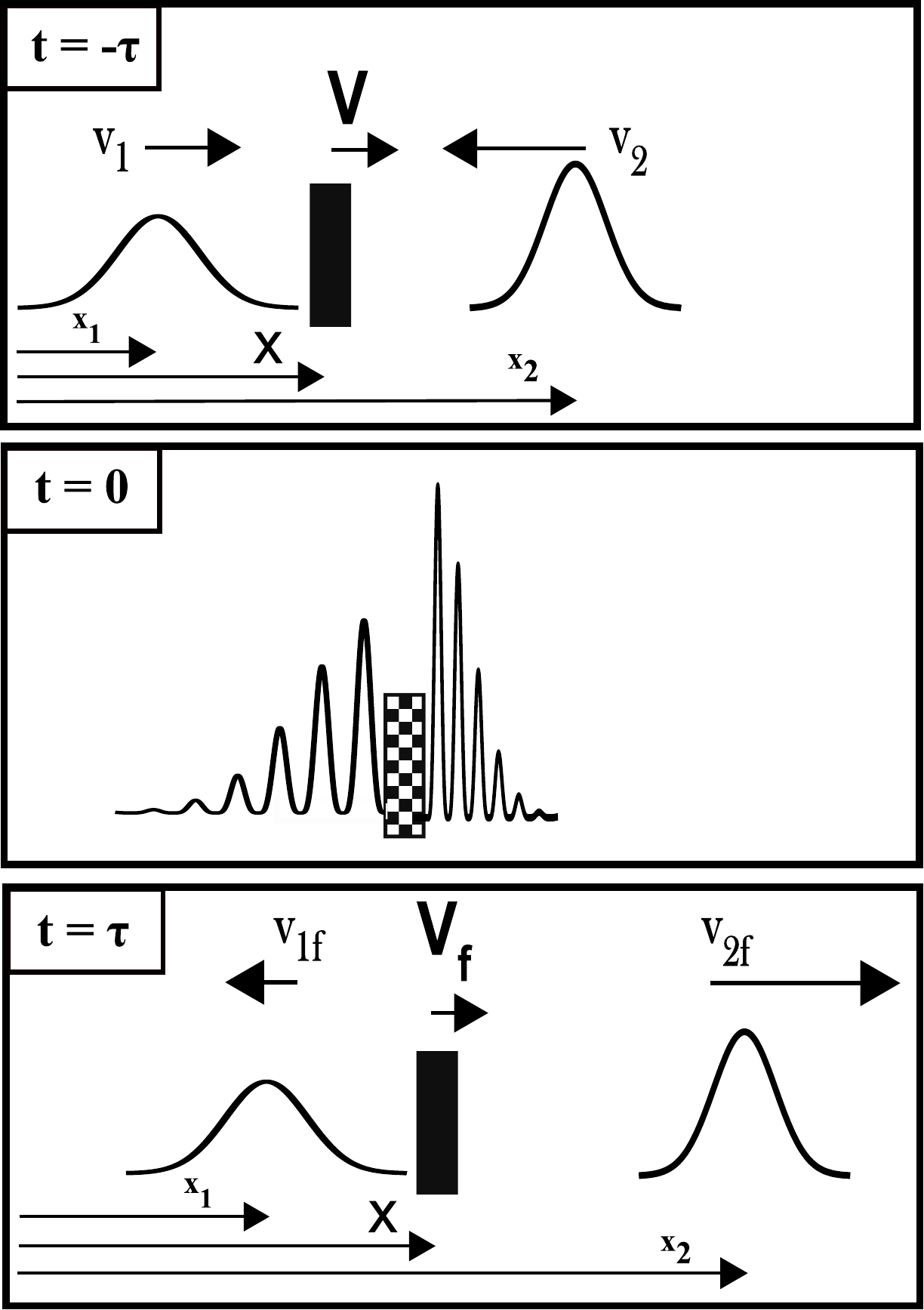}
\caption{PDF schematic of the three-body interaction. Type B correlated interference is illustrated at $t=0$.}
\label{fig:overview}
\end{figure}

Fig. \ref{fig:overview} shows a schematic representation of the PDF for a three-body interaction assuming simultaneous measurement of all three bodies. Actual PDFs are given later. The particle substate PDFs before interaction are the Gaussians at $t=-\tau$, one of which, to the left of the ``mirror", is moving to the right, while the other, to the right of the ``mirror", is moving to the left. The ``mirror" is represented as the black rectangle, moving to the right. These are referred to as particle $1$, the mirror, and particle $2$, with masses $m_{1}$, $M$, $m_{2}$, initial velocities $v_{1}$, ${\textsf V}$, $v_{2}$ (shown as negative in this figure), and coordinates $x_{1}$, {\textsf X}, $x_{2}$, respectively. Although the particles are shown on opposite sides of the mirror, the formalism described below allows for both particles to be on either side. It is assumed that they interact only with the mirror.

At $t=0$ interference between the incident and reflected particle substates is shown as oscillations in the PDF, while interference between the mirror substates, which have and have not reflected the particles, is represented by the rectangular checkerboard pattern rather than the solid rectangle. This snapshot at $t=0$ illustrates a region of correlated interference, although the correlation cannot be shown in such a simple schematic. The figure also cannot illustrate the complexity of interference patterns generated by wavegroup substates of differing sizes which are manifest as correlations in the coincidence rates for the particles and mirror measured at different positions. 

At time $t=\tau$ the three wavegroups have separated so that no overlap occurs. The expectation values of the position of each wavegroup substate then obey the laws of classical reflection.

The three-body PDF sketched in fig. \ref{fig:overview} at $t=0$ involves a superposition of five states (described in sec. \ref{sec:theory}), two examples of which are particle $1$ reflecting before $2$ and particle $2$ reflecting before $1$. For a large number of N distinct particles reflecting from the mirror this number of states scales as N! since the reflected energies and momenta of all the particles and mirror differ depending on the order in which reflection occurs (superposition in a quantum computer involves $2^N$ states with N qbits). 

\subsection{Comparison: two-body double slit and reflection interferometers}
\label{sec:compareInt}

Consider the similarities between two-body correlation interferometry for only one microscopic particle reflecting from a mesoscopic mirror and a {\em two-body} quantum treatment of the center of mass motions of a microscopic particle and a mesoscopic double slit through which the particle travels. The double slit is constrained to move only along the direction of a line drawn between the slits. These are examples of type B systems in which the particle and mirror or particle and double slit are initially uncorrelated. Their interaction, however, results in correlated interference.

\begin{center}
\begin{figure}
\includegraphics[scale=0.30]{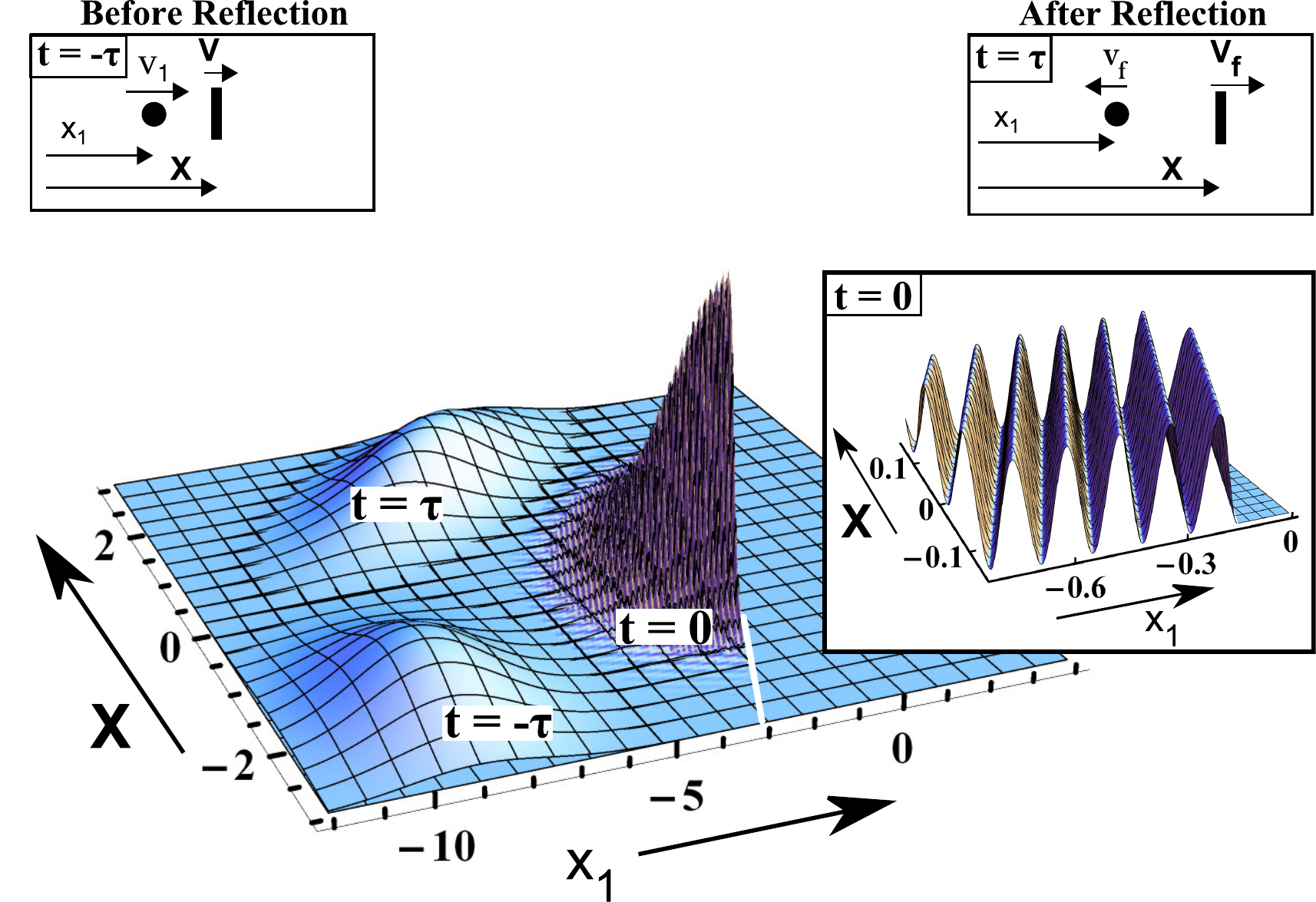}
\caption{Three sequential two-body joint probability density snapshots vs coordinates $(x_{1},{\textsf X})$ for only reflection of particle $1$ from the mirror. The lower PDF waveform moves toward the diagonal white line, corresponding to $x_{1}={\textsf {\textsf X}}$, then reflects in the middle snapshot where the incident and reflected two-body wavefunctions ``overlap", and finally it moves away from the diagonal in the upper snapshot. The upper left inset fig. is a schematic of the ``classical" analog before reflection while the upper right inset fig. is that after reflection. The middle right inset is a magnified version of the $t=0$ plot. There is no classical analog for the middle snapshot. Parameters are: $M/m_{1}=100$, $V/v_{1}=0.2$, and $\Delta V/ \Delta v_{1} = 0.02$}
\label{fig:2body}
\end{figure}
\end{center}

To gain familiarity with such systems, a two-body PDF for reflection of {\em only} particle $1$ from the mirror is shown in fig. \ref{fig:2body}. This is represented schematically in fig. \ref{fig:overview} by eliminating particle $2$.

Three cases of interest are (A) both the particle and mirror or particle and double slit are in localized states \cite{localized}, (B) both are in non-localized states, (C) the mirror or double slit is in a localized state while the particle is not.

For (A) no interference is exhibited since it is possible to determine if the particle has or has not reflected or gone through one or the other slit simply from the two-body wavegroup location along the particle-mirror or particle-double slit coordinates (the centers of masses of the particle and slits then behave as classical objects). Fig. \ref{fig:2body} would then consist of a PDF similar to a delta function replacing the Gaussian distribution shown with no interference at $t=0$.

For (B) a correlated fringe pattern exists along the particle-mirror coordinates, similar to that shown in the middle right inset of fig. \ref{fig:2body}, or along the particle-double slit coordinates. Destructive interference, in the two-body formalism, then correlates the inability to measure the particle and the mirror or particle and double slit at certain locations.  

To illustrate case (C) let the position of the localized mirror be at ${\textsf X}=0$ in fig. \ref{fig:2body}. The interferometric oscillation in the PDF, shown in fig. \ref{fig:2body} at $t=0$, is then concentrated in the $(x_{1},{\textsf X})$ plane along the $x_{1}$ axis as a slice at fixed ${\textsf X}=0$ \cite{kowalski1a} (yielding a PDF similar to that shown in fig. \ref{fig:macromirror}c). In this case, the marginal PDF associated with measuring only the particle involves integration of this two-body PDF along the mirror coordinate. The result is the expected one-body standing wave PDF for a non-localized particle reflecting from a stationary classical mirror. A similar integration along the double slit coordinate yields the familiar double slit interference pattern. 

This marginal PDF also conceals the two-body correlation which remains. For example, consider a point in the $(x_{1},{\textsf X})$ plane in the $t=0$ inset of fig. \ref{fig:2body} (or fig. \ref{fig:macromirror}c) corresponding to destructive interference. The particle and mirror are correlated in that neither's center of mass will be measured at this position. Yet such correlation is revealed only in a measurement of both the particle and mirror and not in the marginal PDF. However, for a classical mirror or double slit no such interferometric correlation exists in principle since these classical bodies cannot be in a superposition state of having and not having reflected the particle or of recoiling from the particle which traverses both the upper and lower slits. 

One of the more unusual aspect of quantum many-body systems is the existence of correlated interference while there is no interference in the associated one-body marginal result. This is a consequence of the marginal PDF concealing the existing correlated interference. An illustration, in the context of case (B), involves measuring only the reflecting particle for the two-body correlated interference PDF shown in the $t=0$ inset of fig. \ref{fig:2body}. Integrating this PDF over the mirror coordinate, to obtain the marginal PDF, results in the reflecting particle fringes ``washing out." Nevertheless, correlated interference remains if both the particle and mirror are measured. Measuring {\em only} the particle after it has traversed a double slit will similarly hide the existing correlated interference if the double slit and particle are both in non-localized states \cite{doubleslit}. 

Again, no such correlated interference exists in principle if the mirror or double slit behave as classical objects, which cannot be in the required superposition states. In this case, only one-body interference is present.

\subsection{Wavefunction collapse in a type B many-body interferometer}
\label{sec:collapse}

Some types of measurements are possible only in many-body systems. One category involves correlation but not interferometry. For example, the momentum of one body and the position of the other can be measured at $t=\tau$ in fig. \ref{fig:2body}. Another category is of measurements on a many-body wavegroup involving type B correlated interference, such as shown at $t=0$ in fig. \ref{fig:overview}. These can be done asynchronously (as discussed in subsection \ref{sec:asynch}). Or, the mirror in fig. \ref{fig:overview} can be replaced with a component which is a beamsplitter for particle $1$ and a mirror for particle $2$. Measurement of a non-local particle $1$ retro-reflecting from the beamsplitter yields three-body correlated interference. Measurement of it transmitting results in only two-body correlated interference between particle $1$ and the mirror. In this and in an asynchronous many-body interferometer, measurement modifies interference rather than destroying it. This is discussed in more detail in sections \ref{sec:asynch} and \ref{sec:beamsplitter/mirror}.

To illustrate related issues, let the {\em non-local} incident three-body state for fig. \ref{fig:overview}, which has a well defined energy and momentum but is not shown in this figure, be $\Psi_{{\bf 1}}$ while the {\em non-local} state associated with a particular reflection order be $\Psi_{{\bf i}}$. These incident and reflected states have the same total energy and momentum. That is, the sum of the frequencies or wavevectors of the particles and mirror substates for any $\Psi_{{\bf i}}$ is equal to those sums for $\Psi_{{\bf 1}}$. Since the reflection order and therefore path information is not known, the system is described by $\Psi_{{\bf tot}}=\sum_{i=1}^{n}\Psi_{{\bf i}}$ (which is calculated in detail below).

Measurements are then made on an ensemble of these three-body systems. Let a mirror momentum measurement on one of the members of this ensemble, with sufficient resolution to distinguish the reflection order, result in a particular value associated with $\Psi_{{3}}$. A momentum measurement on the two particles of different masses will then yield values commensurate with conservation of momentum for $\Psi_{{3}}$. Therefore, $\Psi_{{\bf tot}}$ ``collapses" into $\Psi_{{3}}$. If N particles of different masses reflect from the mirror then the {\em non-local} many body superposition state similarly ``collapses'' when either the momentum of the mirror or the momentum of only one of the N particles is measured with the appropriate resolution.

However, the mirror momentum measurement may have insufficient resolution to determine the reflection order. The unmeasured bodies in this many-body system then remain in a superposition state. An example of measurement with limited resolution is found in the two-body localized double slit traversed by a non-local particle. A momentum measurement of the double slit, which does not have the resolution to determine through which slit the particle traverses, does not eliminate the particle interference. The particle then remains in a superposition state after such a measurement.

However, {\em position} measurements on such a non-local three-body system do not yield the order of reflection since the positions of the bodies are indeterminate. The particle, measured at a given location, could have come from the incident or any of the reflected states. Correlated interference is then revealed as measurements of particle and mirror {\em positions} are made on many members of the ensemble. The duality between momentum and position in quantum mechanics is then manifest in momentum and position measurements in the following manner: (1) on a {\em non-local} particle-mirror system a momentum measurement reveals the order of reflection while a position measurement does not (2) on a {\em local} particle-mirror system a position measurement reveals the order of reflection while a momentum measurement does not.

\subsection{Correlated interference: closed and open interferometers}
\label{sec:closedopen}

The interference shown in fig. \ref{fig:2body} can be contrasted with that from a two-body quantum treatment of a particle interacting with a finite potential well for one dimensional center of mass motion of both the particle and finite well (replace the center of mass of the mirror in fig. \ref{fig:2body} with the center of mass of the well) \cite{well}. An incident particle wavegroup substate with a size much larger than that of the well and interacting with a localized well yields results which are related to reflection of a pulse of light from a thin film. For example, interference for the reflected two-body wavegroup, determined by the well depth and width, persists long after the wavegroup has left this interaction region. It essentially encompasses the whole two-body wavegroup envelope rather than generating fringes which vary with position within it. If the two-body wavegroup shown in fig. \ref{fig:2body} at $t=\tau$ had reflected from a well then such particle-well interference would be manifest in a variations of this wavegroup envelope similar to variations of a light pulse envelope reflecting from a thin film.

In this case, correlated destructive interference of the wavegroup in reflection corresponds to not being able to measure the particle and well in positions of having reflected and recoiled, respectively. The particle then must have transmitted through the well. This example illustrates the need for correlation in the interference between the substates of a many-body system to preserve conservation of energy and momentum. 

It is also an example of correlated interference in a {\em closed-interferometer}. The interference associated with the superposition imposed by such an interferometer is fixed beyond the output port and therefore is independent of the location of the detectors which measure the particle and well. Other examples involve fixed path length Mach-Zehnder, Talbot, and Michelson interferometers \cite{closed int}. 

On the other hand, fig. \ref{fig:2body} illustrates perhaps the simplest {\em open-interferometer}, where the interference fringes vary with position, in this case within the wavegroup envelope. Another {\em open-interferometer} example is that of two laser beams crossing where the fringes are localized to the intersection region of the beams \cite{beams}.

The closed-open interferometric dichotomy is most obvious for non-local states, in which case, the PDF beyond the output ports for a {\em closed-interferometer} is independent of the coordinates of the bodies. It is a function only of the fixed path differences within the interferometer. In an {\em open-interferometer} this PDF depends on the coordinates of the bodies.

A wavegroup adds a constraint on this ``open" interference. For example, the fringes shown in fig. \ref{fig:2body} at $t=0$ would not disappear at $t=\tau$ for a non-local state. Yet they do for the wavegroup. The interference fringes are washed out at $t=\tau$ due the position dependent phase shifts in the superposition of states with different wavevectors which comprise the wavegroup. It is then possible to determine if the particle did or did not reflect simply from a measurement of the position of the two-body wavegroup at $t=\tau$ in fig. \ref{fig:2body}. Such a determination is not possible for a non-local state in this case, yielding open interferometer fringes even when $t=\tau$. 

A two-body system can have a pair, one for each body, of either closed or open substate interferometers. For example, an open-closed interferometer has a PDF which depends only on the coordinates of the body corresponding to the open interferometer. A three-body system could have a permutation of each interferometer type for each particle. Figs. \ref{fig:2body} and \ref{fig:overview} illustrate open-open and open-open-open systems respectively. Examples of closed-closed systems are given in the next section.

Reflection of the particle-well wavegroup has both transient open-open and persistent closed-closed interference properties. During the initial interaction near $t=0$ the incident and reflected wavegroups form fringes with spatial dependence. However, after this overlap disappears, only the {\em closed-closed interferometer} nature of the particle-well interaction remains. The resulting persistent interference is manifest in the magnitude of the wavegroup envelope at $t=\tau$ varying with finite well spacing (e.g. the two-body reflected particle-well wavegroup would essentially disappear for destructive interference). Such persistent interference is a consequence of it being impossible for any measurement of the particle and/or well to result in path information about the particle having reflected from one well boundary or the other, whereas the transient interference involves not knowing if the center of masses of the particle and well did or did not reflect.

While measurement of quantum correlation is difficult, its verification in an open interferometer is even more challenging, particularly when the detector must measure the particle at a position within a small fringe spacing. To mitigate these experimental difficulties, designs which transform open to closed interferometers are discussed in sec. \ref{sec:beamsplitter/mirror}. 

Experimental evidence of such coupled interferometers in two-body microscopic systems is considered next. The following examples treat most interferometer components as classical potentials while the correlation is generated via recoil between an atom and the photon it emits or scatters either before or within the interferometer.

\subsection{Experimental evidence for correlated interference in recoil}
\label{sec:literature}

The first example involves a photon emitted by an atom \cite{tomkovic}. The atom initially moves in a direction perpendicular to that of the emitted photon. A {\em closed-interferometer} is constructed for the photon using only a retro-reflecting mirror (a classical potential) which superposes photon paths associated with its emission in opposite directions from the atom. The atom does not simultaneously emit two photons, each in opposite directions. Rather, it is not known in which direction the one photon is emitted. The superposed photon states move in the same direction since one is retro-reflected. They have the same momentum, having each recoiled in emission.

A closed interferometer is also constructed for atom states which have recoiled in opposite directions by using a Bragg grating placed downstream from the emission region to combine these atom states with the same momentum. Correlation between the atom and photon interferometer outputs is a consequence of conservation of energy and momentum in recoil. This is an example of a correlated closed-closed interferometer since the PDF beyond the output ports depends on neither the photon nor atom coordinates.

Interference in the output of the atom interferometer is then correlated with the output of the photon interferometer. In addition it is correlated with the path difference in the photon interferometer. However, rather than measuring correlated interference between the outputs of the atom and photon interferometers, a marginal PDF involving measurement of only the atom interferometer output was obtained. Nevertheless, the correlation between these interferometers is manifest as changes in the fringe pattern in the output of the atom interferometer when the path difference in only the photon interferometer is varied. A similar result is derived in sec. \ref{sec:coupled} for one particle reflecting from a mirror.

The second example, related to two-body correlated interference in reflection, involves the scattering of a photon from an atom while the atom traverses a MZ (Mach-Zehnder) interferometer \cite{chapman}. The photon ``reflects" from the atom while it is in both arms yielding interference similar to that from two point sources that are separated by the distance between the atom states in the MZ arms. The scattered photon momentum states are in general not the same, resulting in the open nature of the interference. The atom recoil states similarly differ yielding an open atom interferometer substate.

As in the previous example, a marginal PDF involving measurement only at the output port of the atom substate interferometer was obtained. A closed photon interferometer was then constructed by restricting the measurement of the atom to positions at its output port corresponding to interference of both the photon states in the same direction. This then results in a superposition of both photon states with the same momentum and atom states with the same momentum. Inference of correlation between these closed-closed interferometers was then made. However, when not restricting the measurement of the atom in such a manner, the interferometer was open-open. The marginal interference of the atom then vanished as expected. 

The first example discussed above is of a type A system where two particles are initially correlated in a momentum-conserving decay and then each traverses different interferometers whose outputs combine the same momentum photon and the same momentum atom two-body states. This differs from the type A open-open system described by Gottfried \cite{Gottfried} but apart from their open/closed nature they are similar in their correlation due to conservation of energy and momentum in a decay and then the subsequent particle traversal of separate interferometers \cite{Gottfried}. 

The second example discussed above is of a type B system where an initially uncorrelated photon and atom become correlated in their interaction within a MZ interferometer. However, the interaction generating the atom-photon correlation is only between these microscopic particles and not between the particles and other interferometer components. The data can then be interpreted as supporting the treatment of the components or the MZ as classical potentials.

Both of these examples measure interference in closed-closed interferometers which are correlated via recoil of two microscopic bodies. The type B correlated interferometry in reflection shown in fig. \ref{fig:overview} differs from these examples in that the superposition is of substates whose momenta differ in magnitude and direction, in which case the interference is fundamentally that of an {\em open-open-open interferometer}. Another difference is that only the atom and photon are treated as quantum objects in the above examples while the mirrors and gratings of these interferometers are classical potentials. On the other hand, in fig. \ref{fig:overview} all the interferometer components  are treated quantum mechanically.

Yet another difference is that the displacement of the atoms is mesoscopic while the displacement of the mirror is microscopic for a microscopic particle reflecting from a mesoscopic mirror. This large displacement of the atoms, either from recoil in emission or due to traversing a MZ interferometer, facilitates studies which support the model that quantum decoherence is a function of this displacement, as discussed in sec. \ref{sec:environment} \cite{Hackermuller,hornberger}. The prediction that environmental decoherence is mitigated in correlation interferometry with a microscopic particle reflecting from a mesoscopic mirror is then based on the experimental evidence from such atom interferometers.

\subsection{Simultaneous vs. asynchronous measurement}
\label{sec:asynch}

An open substate interferometer introduces a feature present only in quantum many-body systems. As an example, consider the open-open system shown in fig. \ref{fig:2body}. Let the particle be measured first at some location within the particle-mirror correlated interference region of fig. \ref{fig:2body} at $t=0$ while the mirror is measured later. Since both bodies are eventually measured the result is not given by a marginal PDF.

It is not possible to determine if the particle did or did not reflect from the mirror in this measurement. Therefore, the mirror must remain in a superposition state of having and not having reflected the particle \cite{mixture}. This is manifest, after measurement of the particle, as a one-body superposition of mirror states which have and have not recoiled. These one-body mirror states can generate persistent interference fringes beyond the $t=\tau$ snapshot of fig. \ref{fig:2body} \cite{kowalski1}.

This is similar to the two-body system in which the resolution of the double slit momentum measurement is insufficient to determine the particle path through the slits, as was discussed in sec. \ref{sec:collapse}. In both this and asynchronous measurement, the unmeasured body remains in a superposition state after a measurement is made on the other body, which has insufficient resolution to determine path information.

However, a simultaneous measurement of the positions of the particle and mirror, long after overlap of the incident and reflected two-body wavefunctions, does not exhibit any interference, as shown in the $t=\tau$ waveform of fig. \ref{fig:2body}. For the particle and mirror to be measured simultaneously at these positions when $t=\tau$, with uncertainty in their locations constrained by the size of the two-body wavegroup, it is clear that reflection must have occurred, thereby yielding path information and eliminating interference. Yet correlation remains. The location of the reflected particle correlates with that of the recoiled mirror, within the uncertainty associated with the wavegroup size, to maintain conservation of energy and momentum. 

Consider now asynchronous measurement in the persistent interference region using the particle-well system discussed in sec. \ref{sec:closedopen} as an example. After measurement of the reflected particle but before measurement of the well, the well is in a one-body superposition state of having reflected the particle from both of its well boundaries. The one-body well wavefunction, generated after the particle has been measured, is then constructed from the two-body wavefunction by fixing the time and position coordinates of the particle where and when it was measured. 

The experimental advantage of asynchronous measurement in the particle-well system is that the particle can be measured at any reflected position (beyond the transient interference region) while maintaining the one-body well superposition state since the particle interference is not spatially localized, as it is for the interferometer shown in fig. \ref{fig:2body}. Nevertheless, the phases of the resulting one-body well states are affected by the location of the particle measurement since the time and position coordinates of this particle are thereafter fixed in the two-body wavefunction which is used to construct the one-body well state after measurement of the particle.

For the three body case shown in fig. \ref{fig:overview} asynchronous measurement could involve first measuring the position of one particle in the interference region at $t=0$. Mathematically, the time and position coordinates of this particle are thereafter fixed in the three-body wavefunction, which becomes the two-body wavefunction (given as a function of the unmeasured particle and mirror coordinates) that then continues to time evolve. Conceptually, the mirror and other particle remain in a superposition state of the measured particle having and not having reflected. This results in correlated interference between the unmeasured particle and mirror after the first particle was measured. Let the second particle be measured next while the mirror is never measured. The result is then predicted by a marginal PDF of the two-body state which resulted from asynchronous measurement of the initial three-body system. While the calculations in sections \ref{sec:theory} and \ref{sec:wavegroups} correspond to PDFs for synchronous measurements, it is not difficult to modify them to predict both asynchronous and asynchronous marginal effects. However, that is beyond the scope of this already lengthy discourse.

\subsection{Synopsis of the calculations}
\label{sec:synopsis}

An open interferometer in which massive particles reflect from a mirror is the focus of the initial calculations. After deriving the full many-body interferometric correlation, the discussion then turns to indirect measurements on this system. However, correlation in an open interferometer is difficult to measure. Methods to transform open to closed interferometers to mitigate this issue are the focus of the remaining discussion, where more practical methods to measure correlated interference in closed interferometers utilizing photons reflecting from a mesoscopic body are considered. 

The mirror is modeled as a moving delta function potential \cite{kowalski3} where reflection is assumed to occur at its center of mass, with the boundary condition that the wavefunction (incident and reflected three-body states) vanish at $x_{1}=\;${\textsf X~}$=x_{2}$. More realistic boundary conditions limit the generic nature of the results. Only the center of mass motions for the particles and mirror are calculated and the velocities used are assumed to be those which result in reflection of both particles from the mirror.

\section{Three-body reflection in an open interferometer}
\label{sec:theory}

\subsection{Energy eigenstates}

\subsubsection{Incident}
\label{sec:fundamentals}

Assume that the two particles and mirror are initially in an uncorrelated eigenstate of energy before reflection. The separable solution to the Schr\"odinger equation for this non-interacting ``particle-particle-mirror'' state is

\begin{multline}
\Psi_{{\bf 1}}=\psi[x_{1},t] \psi[{\textsf X~},t] \psi[x_{2},t]\propto \\ \exp[i (k_{1} x_{1}-\frac{\hbar k_{1}^{2}}{2m_{1}}t +K {\textsf X~}-\frac{\hbar K^{2}}{2M}t+k_{2} x_{2}-\frac{\hbar k_{2}^{2}}{2m_{2}}t)],
\label{eq:state1}
\end{multline}

where $x_{1}$, $x_{2}$, and ${\textsf X~}$ are the positions along the x-axis of the two ``particles'' and ``mirror'' respectively, while $k_{1}$, $k_{2}$ and $K$ are the respective {\em incident} wavevectors; $k_{1}=m_{1} v_{1}/\hbar$, $k_{2}=m_{2} v_{2}/\hbar$ and $K=M V/\hbar$ with masses $m_{1}$, $m_{1}$, and $M$, and initial velocities $v_{1}$, $v_{2}$ and $V$, as defined in sec. \ref{sec:interferometer}. The PDF for such non-local uncorrelated particle-mirror states then leads to predictions about the probability of simultaneously finding particle $1$ at $x_{1}$, particle $2$ at $x_{2}$, and the mirror at ${\textsf X~}$.

Consider next the energy eigenstates for the system shown in fig. \ref{fig:overview}. These three bodies can exist in 6 possible eigenstates: {\bf (1)} uncorrelated incident, {\bf (2)} particle $1$ has reflected but not particle $2$, {\bf (3)} particle $2$ has reflected but not particle $1$, {\bf (4)} particle $1$ reflected first followed by reflection of particle $2$, {\bf (5)} particle $2$ reflected first followed by reflection of particle $1$, and {\bf (6)} simultaneously reflection of the three bodies. Since it is not known where the bodies are located for such non-local states, the amplitudes for these six ``paths" must be summed.

Interestingly, conservation of energy and momentum do not determine a unique solution for simultaneous three-body elastic reflection \cite{smith}. It is assumed here that the probability for such a collision is small compared with that of the other five ``paths'' and therefore is neglected \cite{comment}. Experimental confirmation of the resulting predictions will validate this assumption.

\subsubsection{Reflected eigenstates {\bf (2)} and {\bf (3)}}
\label{sec:twobody}

The Schr\"odinger equation for eigenstate {\bf 2} is given by the product of a two-body correlated state, due to reflection of particle $1$ from the mirror, with the non-interacting substate for particle $2$. These are separable, yielding $\Psi_{{\bf 2}}= \psi[x_{1},{\textsf X~},t] \psi[x_{2},t]$.

The two-body state is a solution to
\begin{multline}
(\hbar^{2} \partial_{x_{1}}^{2}/2m+\hbar^{2} \partial_{{\textsf X~}}^{2}/2M+\beta \delta[x_{1}-{\textsf X~}] \notag
+i \hbar \partial_{t})\psi[x_{1},{\textsf X~},t]=0,
\end{multline}
where square brackets are used to indicate the argument of a function. For simplicity, the mirror reflectivity is assumed to be unity. Therefore, letting $\beta \rightarrow \infty$, results in the boundary condition that the sum of the incident and reflected states vanish at the mirror and no state exists for $x_{1}>{\textsf X~}$ since the particle cannot move through the mirror.  

The solution, $\Psi_{{\bf 2}}= \psi[x_{1},{\textsf X~},t] \psi[x_{2},t]$, is given by \cite{kowalski1}

\begin{multline}
\Psi_{{\bf 2}}\propto\exp[i (k_{{\bf 1r}} x_{1}-\frac{\hbar k_{{\bf 1r}}^{2}}{2m_{1}}t+K_{{\bf 1r}}{\textsf X~}-\frac{\hbar K_{{\bf 1r}}^{2}}{2M}t)] \\ \times \exp[i (k_{2} x_{2}-\frac{\hbar k_{2}^{2}}{2m_{2}}t)],
\label{eq:state2}
\end{multline}
where $ k_{{\bf 1r}}=m_{1} (2MV-Mv_{1}+m_{1}v_{1})/\hbar(M+m_{1})$ and $ K_{{\bf 1r}}=M(MV-m_{1}V+2m_{1}v_{1})/\hbar(M+m_{1})$. These wavevectors are obtained by determining the velocities of the reflected particle and mirror, $v_{{\bf 1r}}$ and $V_{{\bf 1r}}$, from conservation of momentum and energy in the elastic collision and then using them in the relations $ k_{{\bf 1r}}=m_{1}v_{{\bf 1r}}/\hbar$ and $ K_{{\bf 1r}}=MV_{{\bf 1r}}/\hbar$. 
 
The same procedure for $\Psi_{{\bf 3}}=\psi[x_{2},{\textsf X~},t] \psi[x_{1},t]$ yields
\begin{multline}
\Psi_{{\bf 3}}\propto \exp[i (k_{{\bf 2r}} x_{2}-\frac{\hbar k_{{\bf 2r}}^{2}}{2m_{2}}t+K_{{\bf 2r}}{\textsf X~}-\frac{\hbar K_{{\bf 2r}}^{2}}{2M}t)] \\ \times \exp[i (k_{1} x_{1}-\frac{\hbar k_{1}^{2}}{2m_{1}}t)],
\label{eq:state3}
\end{multline}
where $ k_{{\bf 2r}}=m_{2} (2MV-Mv_{2}+m_{2}v_{2})/\hbar(M+m_{2})$ and $ K_{{\bf 2r}}=M(MV-m_{2}V+2m_{2}v_{2})/\hbar(M+m_{2})$.

\subsubsection{Reflected eigenstates {\bf (4)} and {\bf (5)}}
\label{sec:eigenstates4and5}

The state for particle $1$ reflecting before particle $2$, after both have reflected, is given by
\begin{multline}
\Psi_{{\bf 4}}\propto \exp[i (k_{{\bf 1r}} x_{1}-\frac{\hbar k_{{\bf 1r}}^{2}}{2m_{1}}t+ K_{{\bf 12}} {\textsf X~} \\ -\frac{\hbar K_{{\bf 12}}^{2}}{2M}t+k_{{\bf 2r12}} x_{2}-\frac{\hbar k_{{\bf 2r12}}^{2}}{2m_{2}}t)] ,
\label{eq:state4}
\end{multline}
where the reflected wavevectors are given by $k_{{\bf 2r12}}=m_{2}v_{{\bf 2r12}}/\hbar$, and $K_{{\bf 12}}=MV_{{\bf 12}}/\hbar$ (and $k_{{\bf 1r}}$ is given following eqn. \ref{eq:state2}). The velocities $V_{{\bf 12}}$ and $v_{{\bf 2r12}}$ are of the mirror and particle $2$ after both have reflected in the specified order. Expressions for these velocities again follow directly from conservation of momentum and energy. The reflected wavevectors are given by solving the following eqns: $k_{{\bf 2r12}}\hbar(M+m_{1})(M+m_{2})=2Mm_{2}(MV-m_{1}V+2m_{1}v_{1})+m_{2}(M+m_{1})(m_{2}-M)v_{2}$ and $K_{{\bf 12}}\hbar(M+m_{1})(M+m_{2})=M(M-m_{2})(MV-m_{1}V+2m_{1}v_{1})+2M(M+m_{1})m_{2}v_{2}$.

The state for particle $2$ reflecting before particle $1$, after both have reflected, is given by
\begin{multline}
\Psi_{{\bf 5}}\propto \exp[i (k_{{\bf 2r}} x_{2}-\frac{\hbar k_{{\bf 2r}}^{2}}{2m_{2}}t+ K_{{\bf 21}} {\textsf X~}-\frac{\hbar K_{{\bf 21}}^{2}}{2M}t+ \\ k_{{\bf 1r21}} x_{1}-\frac{\hbar k_{{\bf 1r21}}^{2}}{2m_{1}}t)],
\label{eq:state5}
\end{multline}
where $k_{{\bf 1r21}}=m_{1}v_{{\bf 1r21}}/\hbar$ and $K_{{\bf 21}}=MV_{{\bf 21}}/\hbar$ ($k_{{\bf 2r}}$ is given following eqn. \ref{eq:state3}). The velocities $V_{{\bf 21}}$ and $v_{{\bf 1r21}}$ are of the mirror and particle $1$ after both have reflected in the specified order. Expressions for these velocities again follow directly from conservation of momentum and energy. The reflected wavevectors are given by solving the following eqns: $k_{{\bf 1r21}}\hbar(M+m_{1})(M+m_{2})=m_{1}(M^{2}(2V-v_{1})+m_{1}m_{2}v_{1}+M(m_{1}v_{1}-m_{2}(2V+v_{1}-4v_{2})))$ and $K_{{\bf 12}}\hbar(M+m_{1})(M+m_{2})=M(M^{2}V-M(m_{1}(V-2v_{1})+m_{2}(V-2v_{2}))+m_{1}m_{2}(V+2v_{1}-2v_{2}))$.

\subsection{Three-body eigenstate interference}
\label{sec:interference}

The boundary condition that the three-body energy eigenstate vanish at the mirror, $x_{1}=\;${\textsf X~}$=x_{2}$ is satisfied by the following superposition,
\begin{eqnarray}
\Psi_{{\bf tot}}^{eigenstate}\propto \Psi_{{\bf 1}}-c_{2} \Psi_{{\bf 2}}/2-c_{3} \Psi_{{\bf 3}}/2+c_{4} \Psi_{{\bf 4}}-c_{5} \Psi_{{\bf 5}},
\label{eq:Psitot}
\end{eqnarray}
where the constants $c_{2}$, $c_{3}$, $c_{4}$, and $c_{5}$ are constrained by  $c_{2}+c_{3}-2 c_{4}+2 c_{5}-2=0$. It is assumed that $c_{2}= c_{3}= c_{4}= c_{5}=1$, which results in cancellation of some interference terms. While this choice does not affect the fundamental results it does simplify the calculations presented below. 

Generic results for $PDF_{{\bf tot}}^{eigenstate}=\Psi_{{\bf tot}}^{eigenstate} \Psi^{\ast~ {eigenstate}}_{{\bf tot}}$ using eqns. \ref{eq:state1} $\rightarrow$ \ref{eq:Psitot} are too lengthy to present here. However, for $M\gg (m_{1}, m_{2})$ this reduces to 
\begin{multline}
PDF_{{\bf tot}}[x_{1},{\textsf X},x_{2}]^{eigenstate} \propto \frac{3}{2}-\cos[\alpha] \\ 
+\frac{\cos[\alpha-\beta]}{2}
-\cos[\beta],
\label{eq:EntangledInterference}
\end{multline}
where
\begin{eqnarray}
\alpha=\frac{2m_{1}({\textsf V}-v_{1})(x_{1}-{\textsf X})}{\hbar}~\mbox{and}~\beta=\frac{2m_{2}({\textsf V}-v_{2})(x_{2}-{\textsf X})}{\hbar}.\notag
\label{eq:oneparticle}
\end{eqnarray}

The $\cos[\alpha]$ term is the two-body correlated interference expression or cross term associated with interference between states ${\bf 1}$ and ${\bf 2}$. The $\cos[\beta]$ term is due to interference between states ${\bf 1}$ and ${\bf 3}$. The $\cos[\alpha-\beta]$ term is due to interference between states ${\bf 2}$ and ${\bf 3}$. For $M\gg (m_{1}, m_{2})$ states ${\bf 4}$ and ${\bf 5}$ do not contribute to eqn. \ref{eq:EntangledInterference}. 

One issue which is fundamental in determining the quantum-classical boundary is that of interferometric effects which do not become imperceptible in the limit of large mirror mass. For example, in eqn. \ref{eq:EntangledInterference} the fringe spacing, which is approximately determined by $\alpha$, $\beta$, or their difference cycling through $2 \pi$ radians, is given by displacements along the coordinates which are inversely proportional to the masses of the microscopic particles reflecting from the mesoscopic mirror and {\emph{not}} inversely proportional to the mirror mass, as discussed in sec. \ref{sec:reflection}.

\subsection{Interference of wavegroups}
\label{sec:wavegroups}

Any experimental realization of the interferometer sketched in fig. \ref{fig:overview} will involve wavegroups. The incident wavegroup is assumed to be given by a integral of $\Psi_{{\bf 1}}^{eigenstate}$ over Gaussian distributions of the initial wavevector components $k_{1}$, $k_{2}$, and $K$. The mirrors amplitude is proportional to $\exp [-(K-K_{0})^{2}]/(2 \Delta K^{2})$ where the peak of the distribution is at $K_{0}$ and $\Delta K$ is its width, while for particle $1$ the amplitude is proportional to $\exp [-(k_{1}-k_{10})^{2}]/(2 \Delta k_{1}^{2})$, where the peak of the distribution is at $k_{10}$ and $\Delta k_{1}$ is its width. A similar expression is used for particle $2$. In the same manner, an analytic expression for the incident and reflected wavegroups is obtained from $\Psi_{{\bf tot}}^{eigenstate}$ yielding the three-body wavegroup $\Psi_{{\bf tot}}^{wavegroup}$. Analytic expressions are obtained for these integrals. 

For particle masses much less than that of the mirror these wavegroup calculations produce fringe spacings which do not perceptibly decrease with increasing mirror mass as described in section \ref{sec:reflection}. The wavegroup fringe pattern shown in fig. \ref{fig:2body} is essentially that of the non-local calculation modified by an envelope factor associated with the two-body wavegroup.

These Gaussian wavegroups yield incident and reflected peaks of the three-body wavegroup substates which overlap at $t=0$ as shown schematically in fig. \ref{fig:overview}. For localized particles the reflection order is associated with that of the reflection of these peaks. However, as the widths of the Gaussian wavegroup substates increase and begin to overlap the reflection order then becomes indeterminate. The peaks can reflect in one order while the energy and momentum transfered to the particles corresponds to a different reflection order. As these wavegroup widths increase and overlap there is a transition from particle to wave behavior which is manifest as interference. However, the resulting correlated interference is nevertheless modulated by the envelope of these Gaussian wavegroup substates. Modifications of the results above to include delays in the reflection of the peak of one particle wavegroup with respect to the other are discussed in the appendix \ref{sec:appendixa}.

It is difficult to display three-body PDF plots of correlated interference, as shown schematically in fig. \ref{fig:overview}. Instead, these plots are shown as a function of only the coordinates of particle $1$ and the mirror (a slice in the $(x_{1},{\textsf X},x_{2})$ space), for a position of particle $2$ that is located within the interferometric PDF oscillations sketched in fig. \ref{fig:overview}.

Another simplification, to illustrate such three-body PDFs, incorporates a broadband mirror substate wavegroup (narrow spatial width compared with the fringe structure) with narrow band (large spatial width compared with the fringe structure) particle substate wavegroups. This transforms the wavegroup shown in fig. \ref{fig:2body} into one with an elliptical footprint in the $(x_{1},{\textsf X})$ plane, whose large major and small minor axes are along $x_{1}$ and $X$, respectively. 

\begin{center}
\begin{figure}
\includegraphics[scale=0.3]{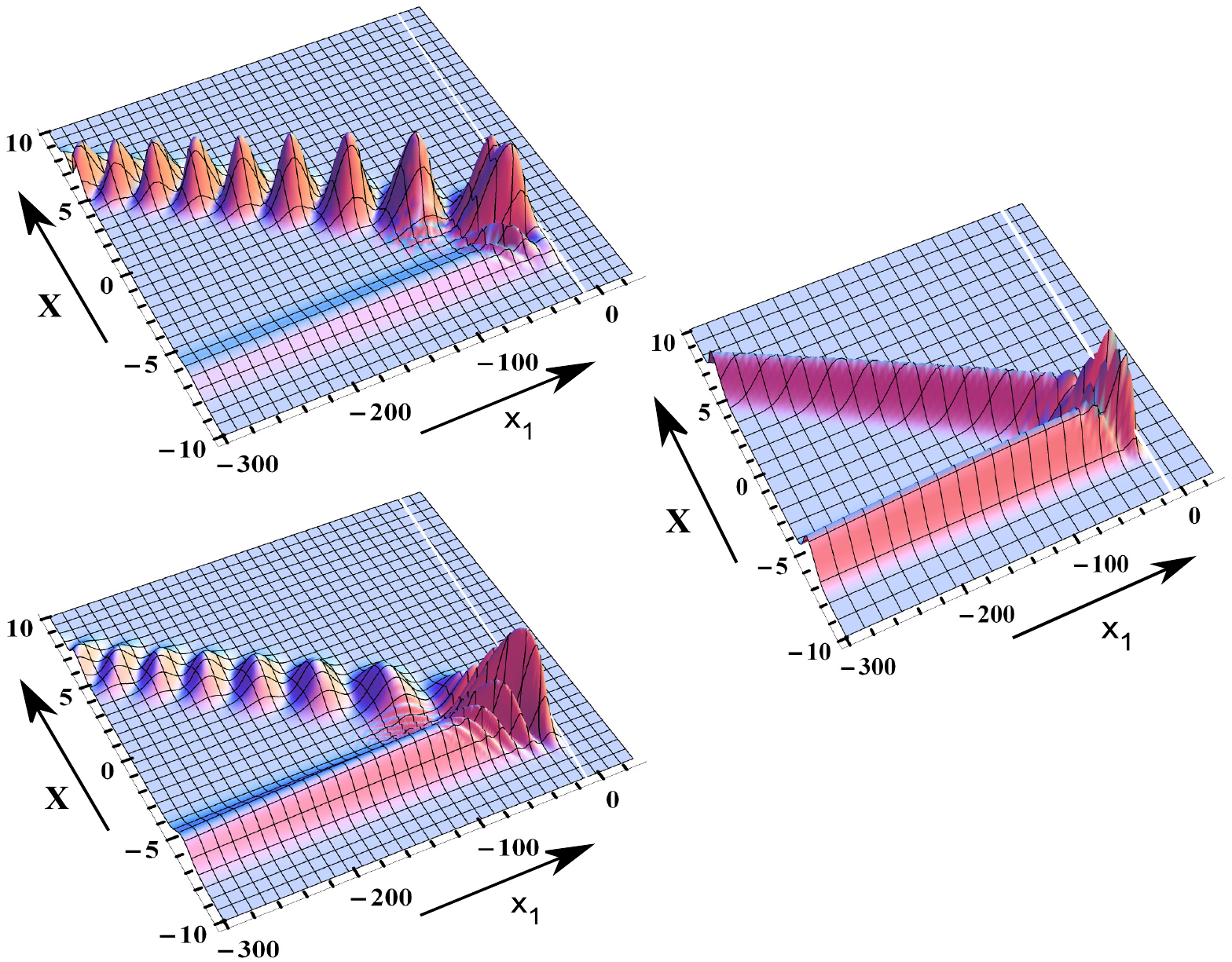}
\caption{A reflecting three-body wavegroup with an elliptical footprint in the $(x_{1},{\textsf X})$ plane. The parameters in the upper and lower left three-body PDF correlated interference plots differ only by the particular value used for $x_{2}$. Using the same parameters, the right plot shows a two-body PDF plot for reflection only of particle $1$ from the mirror.}
\label{fig:3bodyint}
\end{figure}
\end{center}

To illustrate the effect of wavegroup structure on interference compare the right graph shown in fig. \ref{fig:3bodyint} to the snapshot at $t=0$ shown in fig. \ref{fig:2body}, both of which are two-body PDFs. In one case the recoil of the mirror is insufficient to prevent overlap of the incident and reflected two-body wavegroups, yielding the interference in the $(x_{1},{\textsf X})$ plane as shown in fig. \ref{fig:2body}. In the case of the elliptical wavegroup shown in fig. \ref{fig:3bodyint}, the recoil is sufficient to prevent this overlap and therefore prevent interference. A measurement of the particle at a given position $x_{1}$ then yields two mirror positions, those associated with the particle reflecting and not reflecting. The mirror position after recoil, in this case, compared with its initial position is larger that the uncertainty in its position. While interference has disappeared correlation has not. The particle measured at a given position is correlated with the two mirror positions corresponding to the particle having and not having reflected.

The interferometric effects due to the addition of the third body are shown in fig. \ref{fig:3bodyint} for the snapshot at $t=0$. The only difference between the upper and lower left side plots in fig. \ref{fig:3bodyint} is the value of $x_{2}$. However, in both plots, $x_{2}$ is fixed to be within the position of interferometric PDF oscillations sketched in fig. \ref{fig:overview}. The horizontal ridge along the $x_{1}$ axis corresponds to a superposition of paths {\bf 1} and {\bf 3} or interference of states in which particle $1$ has not reflected while the mirror has and has not reflected particle $2$. The diagonal oscillations in the PDF correspond to a superposition of paths {\bf 2}, {\bf 4}, and {\bf 5}, which are defined above. This PDF is a manifestation of the three-body interferometer. Long after the overlap shown in fig. \ref{fig:3bodyint} the elliptical PDF moves away from the white line corresponding to $x_{1}={\textsf X}$ as does the two-body wavegroup shown in fig. \ref{fig:2body} at $t=\tau$.

\begin{center}
\begin{figure}
\includegraphics[scale=0.3]{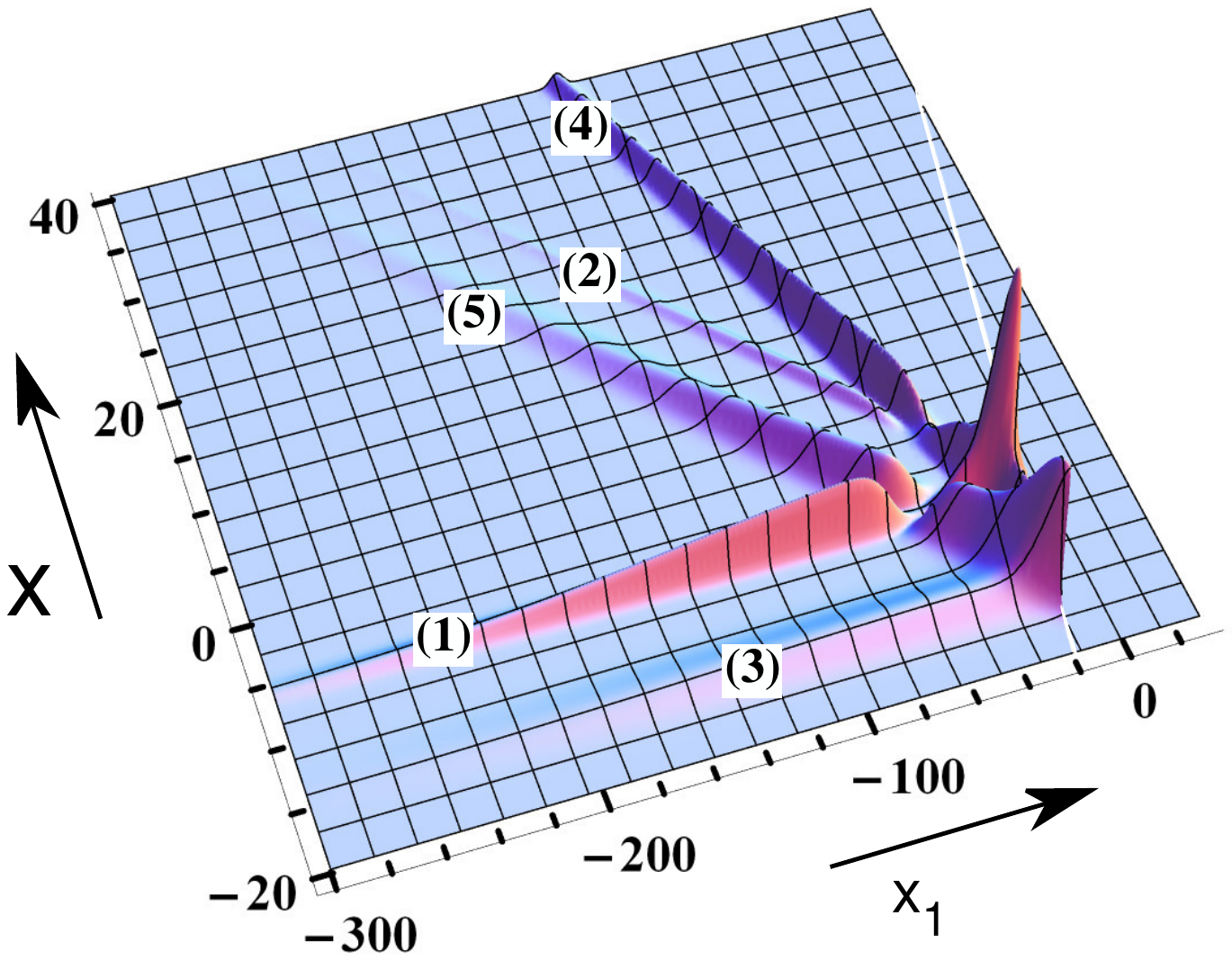}
\caption{Fig. \ref {fig:3bodyint} parameters have been modified to illustrate the system acting as a beamsplitter. The paths associated with the separate ridges are labeled as described in the text. The lack of overlap of these paths prevents the interference between some of these paths which is shown in fig. \ref {fig:3bodyint}.}
\label{fig:beamsplitter}
\end{figure}
\end{center}

Fig. \ref{fig:beamsplitter} is a modification of fig. \ref {fig:3bodyint} which involves narrowing the spatial width of the mirror substate wavegroup while also increasing the mass of particle 2, relative to the parameters used in fig. \ref{fig:3bodyint}. Fig. \ref{fig:beamsplitter} illustrates reflection acting more as a beamsplitter than an interferometer. The paths associated with the separate ridges are labeled as described in sec. \ref{sec:fundamentals}. The reflection of particle $2$ then imparts sufficient momentum to separate these narrower wavegroups, thereby eliminating the interference shown in fig. \ref{fig:3bodyint}. For example, the two interfering states parallel to the $x_{1}$ axis of fig. \ref{fig:3bodyint}, corresponding to paths $\bf{1}$ and $\bf{3}$, which no longer overlap in fig. \ref{fig:beamsplitter}. Since state $\bf{3}$ involves reflection of particle $2$ but not of $1$ its manifestation in these plots is an offset along the ${\textsf X}$ axis due to the impulse delivered to the mirror by particle $2$.

For a measurement of particle $1$ at $x_{1}$ and particle $2$ at the value used in the figure, the mirror is in $5$ distinct positions one along each ridge of the PDF, each associated with a path, and labeled numerically as defined above. Particle $1$, found at such a location, must then be correlated with the mirror located at a position associated with having reflected the particles via  paths $\bf{1}$ through $\bf{4}$.  Although the different paths no longer overlap, the system remains in a superposition of all the paths and not in a separable state. 

Verification of such correlated interference requires simultaneous measurement of both particles and the center of mass of the mirror with instruments having a spatial resolution smaller than the fringe spacing. For a static mesoscopic mirror reflecting microscopic particles, this spacing is about half the deBroglie wavelength of the particles, which at $\sim 1~\mu$m for ultracold atoms \cite{cronin}, potentially satisfies this requirement. Additionally, the effects of longitudinal coherence length, $l_{c}$, need to be considered. For ultracold atoms $l_{c} \leq 10~\mu$m \cite{cronin}, while for a mirror this is given by $l_{c} \approx \lambda^{2}/\Delta \lambda = \lambda {\textsf V}/\Delta {\textsf V}$ \cite{hasselbach}. If the uncertainty in the mirror velocity is determined by its thermal equilibrium with the environment then $\Delta {\textsf V}_{thermal} \approx \sqrt{2k_{B}T/M}$ yielding for the mirror, $l_{c}^{thermal} \approx h/\sqrt{2Mk_{B}T}$. Constraints on indirect measurement of correlated interference involving these coherence lengths are discussed in the next section.

\subsection{Indirect measurement}
\label{sec:marginal}

Measuring interferometric correlations among a subset of bodies is more practical than verifying the full correlation. When applied to the quantum-classical transition such an indirect measurement must reveal the mesoscopic mirror's superposition state via interferometric correlations between only the reflecting particles.

To predict such effects, the many body PDF is integrated over ${\textsf X}$, reducing the coordinate space spanned by the three-body state. This procedure, when applied to the PDF shown in the $t=0$ inset of fig. \ref{fig:2body}, virtually eliminates interference. Yet, measuring particle $1$ for the PDF shown in fig. \ref{fig:3bodyint} while particle $2$ is measured at the location used in this figure yields a marginal PDF which does exhibit three-body interference. These examples illustrate the need for careful experimental design in revealing the quantum behavior of the mirror in an indirect measurement.

Two conditions required to achieve this are first that the mirror must be in a state in which the order of reflection is indeterminate. This is most easily satisfied by reflecting at least one non-local particle. Both states shown in figs. \ref{fig:3bodyint} and \ref{fig:beamsplitter} satisfy this condition. Second, these states must overlap (requiring a small momentum transfer), as illustrated in fig. \ref{fig:3bodyint}, and not simply exhibit the superposition shown in fig. \ref{fig:beamsplitter}. 

Even if the superposed states do not overlap in the $(x_{1},{\textsf X})$ plane, both a small momentum transfer and non-local behavior of particle $2$ may be sufficient to maintain interference in the $(x_{1},{\textsf X})$ plane. An example of this is given in sec.\ref{sec:other marginal}. Determining simple constraints on correlated interference is more difficult for the reflection of particles with such different momentum transfers and non-local properties. In deriving such relationships, described below, it is assumed that each particle is sufficiently non-local and the momentum exchanged in each collision with the mirror is small enough to exhibit interference independent of the other body.

For such overlap of the mirror substates to occur, the mirror coherence length needs to exceed its recoil induced displacement $\Delta D_{3body}$ generated by the particles along the different paths. In so doing, fig. \ref{fig:beamsplitter} is transformed into fig. \ref{fig:3bodyint}. While the coherence length is small for a mesoscopic object, the mirror center of mass displacements for microscopic objects reflecting from a mesoscopic mirror in different orders can be smaller. Assuming a thermal coherence length for the mirror, this is expressed as $l_{c}^{thermal} > \Delta D_{3body}$. 

An estimate of $\Delta D_{3body}$ is next given for the different cross terms in eqn. \ref{eq:EntangledInterference}. Those that do not vanish in the limit of microscopic particles reflecting from a mesoscopic mirror are paths $\bf{1}$, $\bf{2}$, and $\bf{3}$, each of which involves only one-body reflection. For microscopic masses reflecting from a mesoscopic mirror it is shown next that $\Delta D_{3body} \propto m_{1}/M$ or $m_{2}/M$. Since $l_{c}^{thermal} \propto 1/\sqrt{M}$, $\Delta D_{3body}$ decreases faster than $l_{c}^{thermal}$ for increasing $M$.

The positions of the wavegroup substate peaks are approximated using classical kinematics in reflection, where the group velocities and the positions of these wavegroup substate maxima are associated with the velocities and positions of the respective bodies. The differences in mirror positions between these states, denoted in the superscript of $\Delta D$, are then 
$\Delta D_{2body}^{\bf{1}\bf{2}} \approx 2m_{1} (v_{1}-V)t/M$, $\Delta D_{2body}^{\bf{1}\bf{3}} \approx 2m_{2}(v_{2}-V)t/M$, and $\Delta D_{2body}^{\bf{2}\bf{3}} \approx 2 m_{1} (v_{1}-V) t/M- 2 m_{2} (v_{2}-V) t/M$.

However, unlike the reflection shown in fig. \ref{fig:overview}, the wavegroup peaks for each particle do not have to coincide with the mirror wavegroup peak in three body reflection at $t=0$. Modifications of these calculations to include such offsets are discussed in appendix \ref{sec:appendixa}. If such a delayed collision generates an extra displacement which is larger than the mirror coherence length then three-body correlated interference again vanishes. 

The effect of such a delayed reflection on the position of the mirror wavegroup peak can again be estimated by classical kinematics. Consider a delay in the reflection of the particle $2$ substate wavegroup peak by a time $\tau_{0}$ after that of particle $1$. The difference between the mirror positions after only particle $1$ reflects (state $\bf{2}$) and after only particle $2$ (state $\bf{3}$) reflects with a delay $\tau_{0}$ is then $\Delta D_{2body}^{\bf{2}\bf{3}} \approx 2 m_{1} (v_{1}-V) t/M- 2 m_{2} (v_{2}-V) (t-\tau_{0})/M$. An estimate of the conditions needed for an indirect measurement to reveal the superposition state of the mirror is then $l_{c}^{thermal} > \Delta D_{3body}$ for the different mirror displacements given above.

\begin{center}
\begin{figure}
\includegraphics[scale=0.3]{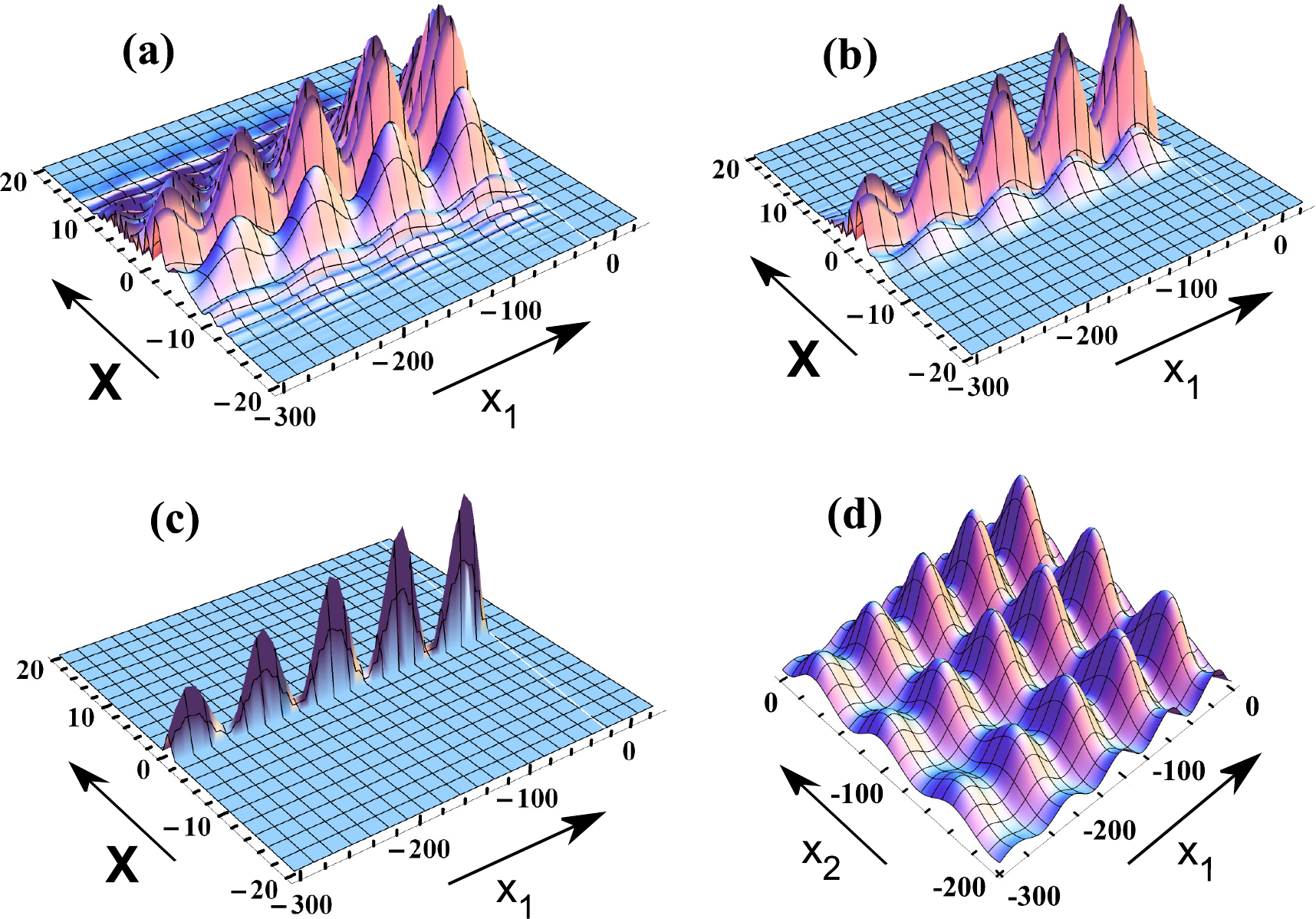}
\caption{Plots (a) through (c) illustrate three-body PDFs with increasing mirror localization, while $x_{3}$ and the size of the particle wavegroups is fixed. Plot (d) illustrates the marginal PDF derived from plot (c).}
\label{fig:macromirror}
\end{figure}
\end{center}

Having satisfied these conditions, correlated interference for wavegroups can then be approximated by energy eigenstate solutions within the overlap region. For example, the locations of the wavegroup fringes shown in fig. \ref{fig:2body} at $t=0$ match well those of calculations which use the superposition of incident and reflected non-local particle-mirror states. The difference predominately involves a factor associated with the Gaussian envelope of the fringes. In addition, the eigenstate approximation to wavegroup interference remains a good approximation even for more localized mirror states as illustrated in fig. \ref{fig:macromirror} (c), which is discussed next. This approximation applied to three-body reflection then utilizes eqn. \ref{eq:EntangledInterference}. 

Another marginal PDF example of a three-body wavegroup system is shown in fig. \ref{fig:macromirror}. Although the parameters are similar to those of fig. \ref{fig:3bodyint}, they are modified to have a much smaller particle $1$ momentum exchange, resulting the incident and reflected states overlapping in a manner similar to that shown in fig. \ref{fig:2body} at $t=0$. In fig. \ref{fig:macromirror} (a) to (c) the width of only the mirror substate is sequentially decreased. These PDFs are determined for a position of particle $2$ that is located within the interferometric PDF oscillations sketched in fig. \ref{fig:overview}. In fig. \ref{fig:macromirror}(a) integration over ${\textsf X}$ essentially washes out any interference in the resulting marginal two-body PDF. However, a similar marginal PDF for fig. \ref{fig:3bodyint} (c) yields the correlated interference two-body PDF shown in fig. \ref{fig:macromirror} (d). Such interference is not possible without the mirror being in a superposition state of different reflection sequences. This example also illustrates the importance of wavegroup structure in the success of having an indirect measurement reveal the quantum behavior of the mirror. 

For the non-local particle substates and localized mirror substate of fig. \ref{fig:macromirror}(c), the marginal PDF can be approximated using eqn. \ref{eq:EntangledInterference}, with the mirror substate being a delta function distribution in position. Integrating over the mirror coordinate then yields 
\begin{multline}
PDF_{{\bf tot}}^{eigenstate}[x_{1},x_{2}] \propto \frac{3}{2}-\cos[\frac{2m_{1}({\textsf V}-v_{1})(x_{1}-{\textsf X}_{0})}{\hbar}] \\ +\frac{1}{2}\cos[\frac{2m_{1}({\textsf V}-v_{1})(x_{1}-{\textsf X}_{0})+2m_{2}({\textsf V}-v_{2})({\textsf X}_{0}-x_{2})}{\hbar}] \\
-\cos[\frac{2m_{2}({\textsf V}- v_{2})({\textsf X}_{0} x_{2})}{\hbar}].
\label{eq:marginalx1x2}
\end{multline}
This differs from the exact wavegroup result shown in fig. \ref{fig:macromirror} (d) by a slight variation in the magnitude of the PDF due to the large but finite Gaussian distributions of the PDFs of the particles. Under these conditions eqn. \ref{eq:marginalx1x2} is a good approximation to the wavegroup interference. 

A comparison of $PDF_{{\bf tot}}^{eigenstate}[x_{1},x_{2}]$ can be made with independent one-body solutions to the Schr\"odinger equation for $M\gg (m_{1}, m_{2})$ while treating the mirror as a classical potential. The PDF for each particle residing on opposite sides of a mirror is,
\begin{multline}
PDF^{classical}[x_{1}]+PDF^{classical}[x_{2}]\approx 2- \\ \cos[\frac{2m_{1}({\textsf V}-v_{1})(x_{1}-{\textsf X}_{0})}{\hbar}]-\cos[\frac{2m_{2}({\textsf V}- v_{2})({\textsf X}_{0}-x_{2})}{\hbar}] \notag.
\end{multline}

This differs from the approximate marginal PDF given in eqn. \ref{eq:marginalx1x2} in not having the $\cos(\alpha-\beta)$ term, which introduces a correlation between a measurement of one particle with that of the other.

An indirect measurement might then consist of ultra cold atoms reflecting from opposite sides of a stationary mirror. Observation of a standing wave atom PDF pattern as predicted in eqn. \ref{eq:marginalx1x2} would then indicate quantum behavior of the mirror without a direct measurement of it.

While the approximation using eqn. \ref{eq:EntangledInterference} is useful in understanding how the interaction of the three bodies is manifest in a marginal PDF, it is incomplete since both the time dependent motion of the wavegroup and its spatial dependence are neglected. It is most accurate during overlap of the three wavegroup substates.

The coordinate space of $PDF_{{\bf tot}}^{eigenstate}[x_{1},x_{2}]$ can again be lowered by reducing the coherence length of particle $2$, resulting in a sequence for fig. \ref{fig:macromirror} (d) which progresses in a similar manner to that of fig. \ref{fig:macromirror} (a) going to fig. \ref{fig:macromirror} (c). The particle $2$ substate then becomes a delta function distribution in position, peaking at $x_{2}=x_{20}$ while that of particle $1$ remains in a broad spatial distribution. This one-body PDF can again be approximated, in the interference region shown in fig. \ref{fig:overview}, by integrating  $PDF_{{\bf tot}}^{eigenstate}[x_{1},x_{2}]$ over $x_{2}$ yielding,

\begin{multline}
PDF_{{\bf tot}}^{eigenstate}[x_{1}] 
\propto \frac{3}{2}-\cos[\frac{2m_{1}({\textsf V}-v_{1})(x_{1}-{\textsf X}_{0})}{\hbar}] \\
+\frac{1}{2} \cos[\frac{2m_{1}({\textsf V}-v_{1})(x_{1}-{\textsf X}_{0})+2m_{2}({\textsf V}-v_{2})({\textsf X}_{0}-x_{20})}{\hbar}]  \\ -\cos[\frac{2m_{2}({\textsf V}- v_{2})({\textsf X}_{0}-x_{20})}{\hbar}].
\label{eq:marginalx1}
\end{multline}

Such a result is useful in estimating the three-body correlation effects in the one-body marginal PDF. However, the approximation again neglects the time evolution and spatial dependence of the wavegroup substates. The validity of this approximation then rests on the order of reflection being indeterminate.  The long duration of the particle $1$ interaction mitigates this constraint.

If, on the other hand, the particle $2$ wavegroup reflects before or after any interaction of particle $1$ with the mesoscopic mirror, then the result is the one-body solution to the Schr\"odinger equation,
\begin{align*}
PDF^{classical}[x_{1}]\approx 1-\cos[\frac{2m_{1}({\textsf V}-v_{1})(x_{1}-{\textsf X}_{0})}{\hbar}]. \notag
\end{align*} 
This again differs from the three-body result given in \ref{eq:marginalx1}.

Evidence of three-body quantum effects can therefore be found in the measurement of the ``standing wave" PDF of only particle $1$. Such an indirect measurement might consist of a nearly static mirror in a system with coherence lengths $l_{c}^{M} \ll l_{c}^{m_{2}} \ll l_{c}^{m_{1}}$. A fixed delay of particle $2$ (constant $\beta$ due to a fixed $x_{2}-{\textsf X}$) is introduced. The interaction time of particle $1$ must be longer than the transit time of particle $2$ to the mirror for the order of reflection to remain indeterminate. Observation of the correlation term in only the standing wave PDF pattern for particle $1$, as a function of $x_{1}-{\textsf X}$ (and therefore $\alpha$), would then indicate quantum behavior of the mirror.

The correlations discussed in this section involve open-open-open interferometers. Measurement of correlated interference in reflection are much easier in a closed interferometer. Methods to convert open to closed interferometers and consequences thereof are discussed in sec. \ref{sec:coupled}.

\subsection{Discussion}
\label{sec:discussion}

The above analysis provides limited insight into the complexity of quantum correlation interferometry in reflection. The following subsections extend that discussion to other cases of particles with differing bandwidths, other marginal PDFs, wavefunction collapse, time dependence, and reflection of zero-rest mass particles.

\subsubsection{Particles with differing bandwidths}
\label{sec:differing bandwidths}

Two-body reflection, with particle and mirror substates which have similar spatial localizations, is illustrated in fig. \ref{fig:2body}. The addition of a second particle, also with similar spatial localization, reflecting as in fig. \ref{fig:overview} and measured at position $x_{2}$ corresponding to the peak position of its Gaussian substate at $t=-\tau,~0,~\tau$, results in a three-body PDF. This PDF, in its $(x_{1},{\textsf X})$ plane, is similar to that shown in fig. \ref{fig:2body}. 

Consider decreasing the bandwidth of only this second particle so that its spatial extent is so large that it now generates substate fringes in reflection along the $x_{2}$ axis in the $(x_{2},{\textsf X})$ plane even for the three times shown in fig. \ref{fig:2body}. This particle is then measured within this fringe region at these three times. The corresponding three-body fig. in the $(x_{1},{\textsf X})$ plane looks similar to that of fig. \ref{fig:2body} with the following modification. A fringe pattern, similar to that shown at $t=0$ in fig. \ref{fig:2body} is now also imposed on the $t=-\tau$ and $t=\tau$ snapshots. There is then interference in $(x_{1},{\textsf X})$ plane where there was none for reflection of a localized second particle. The non-local behavior of this second particle then generates correlated interference even when there is no overlap between the incident and reflected two-body wavegroup substate for particle $1$ and the mirror.

\subsubsection{Other marginal PDFs}
\label{sec:other marginal}

The marginal PDFs, derived in section \ref{sec:marginal}, using the energy eigenstate solution given in eqn. \ref{eq:EntangledInterference}, approximate well the results from the exact three-body solutions for the Gaussian wavegroup described in section \ref{sec:wavegroups}. A discussion of two more marginal PDFs that are also derived from this approximation follows next.

First, let both particles be in non-localized states with a localized mirror. The prediction for measuring only particle $1$ then involves integrating eqn. \ref{eq:marginalx1x2} over the $x_{2}$ coordinate. This yields interference only in the $\cos[\alpha]$ term, with the other terms washing out. That is, the average of these cosine functions in the marginal PDF is zero thereby not reducing the visibility (even when more particles reflect). One-particle marginal interference, while the other particle and the localized mirror are not measured, is therefore preserved even when another non-local particle reflects.

Second, let the mirror be in a non-local state while both particles (or only one) are in localized states. The prediction for measuring only particles $1$ and $2$ then involves integrating eqn. \ref{eq:marginalx1x2} over the ${\textsf X}$ coordinate. This eliminates all interference and is analogous to case (B) described in sec. \ref{sec:compareInt}. To measure one-body particle interference in reflection from this system, while not measuring the mirror, requires that the mirror to be in a localized state.

\subsubsection{Wavefunction collapse with a beamsplitter/mirror}
\label{sec:beamsplitter/mirror}

Next consider replacing the mirror in fig. \ref{fig:overview} with a beamsplitter for particle $1$ while it remains a mirror for particle $2$ \cite{beamsplitter}. Let the incident wavefunction again be a separable Gaussian, $\Psi_{{\bf in}}=\phi[x_{1},t]\chi[{\textsf X~},t]\eta[x_{2},t]$. The total wavefunction for this system is then given by a superposition of three states: a non-interacting state given by $\Psi_{{\bf in}}$, the state where particle $1$ reflects given by $\Psi_{{\bf left}}$, and where it transmits given by $\Psi_{{\bf right}}$. These are expressed as $\Psi_{{\bf left}}=\psi_{ref1}[x_{1},{\textsf X~},x_{2},t] H[{\textsf X}-x_{1}]$ and $\Psi_{{\bf right}}=\phi[x_{1},t]\psi_{ref2}[{\textsf X},x_{2},t] H[x_{1}-{\textsf X}]$, where $H$ is the Heavyside step function. The wavefunctions $\psi_{ref1}[x_{1},{\textsf X},x_{2},t]$ and  $\psi_{ref2}[{\textsf X},x_{2},t]$ are three and two-body correlated reflection states. The resulting ${\bf PDF_{bs}}=\Psi_{{\bf in}}\Psi^{*}_{{\bf in}}+(\Psi_{{\bf left}}\Psi^{*}_{{\bf left}}+\Psi_{{\bf in}}\Psi^{*}_{{\bf left}}+\Psi^{*}_{{\bf in}}\Psi_{{\bf left}}) H[{\textsf X}-x_{1}] +(\Psi_{{\bf right}}\Psi^{*}_{{\bf right}} +\Psi_{{\bf in}}\Psi^{*}_{{\bf right}}+\Psi^{*}_{{\bf in}}\Psi_{{\bf right}}) H[x_{1}-{\textsf X}]$. This is a sum of the incident PDF, a three-body PDF to the left of the beamsplitter, and a two-body PDF to the right of the beamsplitter \cite{kowalski1}.  

This system has the potential for probing constraints on wavefunction collapse, in a manner which is not possible in a one-body formalism. For example, let particle $1$, which interacts with the beamsplitter in the correlated interference region, be absorbed at a particular location to the left of the beamsplitter/mirror by the environment (e.g. bathed by thermal radiation incident perpendicular to the particle $1$ beam in a confined location) but not explicitly measured (e.g. absorbed by a particle counter). The transmitted particle is not absorbed (e.g. it has no such exposure to thermal radiation). If this environmental absorption constitutes a measurement of particle $1$ to the left of the mbeamsplitter/mirror then three-body correlated interference of particle $1$, particle $2$, and the localized bs/mirror will be revealed in measurements on an ensemble of such systems. However, if this particle is not measured by the environment then a marginal ${\bf PDF_{bs}}$ for not measuring particle $1$ (i.e. a marginal PDF of eqn. \ref{eq:marginalx1x2} over $x_{1}$) results in two-body correlated interference involving reflection of particle $2$ from the bs/mirror. Such a method to probe the effect of measurement on interference requires at least a three-body correlated system.

In this case, measurement changes interference while in one body systems measurement only destroys interference. An asynchronous measurement, as discussed in sec. \ref{sec:asynch}, also generates an effect in which interference is modified but not destroyed.

\subsubsection{Time dependence}
\label{sec:time}

The PDF for the energy eigenstate solution given in eqn. \ref{eq:EntangledInterference} has no time dependence. This is a consequence of all five reflecting states having the same total energy. The effect of superposing different total energy eigenstate solutions to form wavegroups is best illustrated in fig. \ref{fig:2body}. The PDF wavegroup envelope travels without fringes until the incident and reflected two-body wavegroups overlap near $t=0$. The fringe pattern remains essentially static as these wavegroup envelopes move across each other. This static fringe pattern disappears when only the reflected wavegroup remains (as shown at $t=\tau$ in fig. \ref{fig:2body}). A similar effect occurs in the three-body interaction.

A classical treatment of a harmonic wave retro-reflecting from a moving mirror, however, involves a time dependent fringe shift of the wave interference. An explanation of how time dependent interference can be introduced into the static correlated quantum PDF, for a particle reflecting from a moving mirror, involves asynchronous measurement \cite{kowalski1}. A similar issue of extracting time dependence from a static entangled state is discussed by Page and Wooters in the context of how time is perceived in the Wheeler-DeWitt equation \cite{page}.

\subsubsection{Correlated interference in reflection with photons}
\label{sec:expt}

Consider replacing the two particles in fig. \ref{fig:overview} with photons, one on each side of the mirror and each with a different energy. The validity of a single photon position dependent wavefunction is controversial \cite{raymer}. However, the experimental data discussed in sec. \ref{sec:literature} support such a correlated interference model in two-body atom-photon recoil. This model is next extended to photon-mirror recoil. Following Fedorov's example of a correlated atom-photon two-body wavefunction \cite{fedorov}, the photon-mirror-photon energy eigenstates {\bf1 } and {\bf 2} are expressed as 
\begin{eqnarray}
\Psi_{{\bf 1}}\propto \exp[i (k_{1} x_{1}-\omega_{1}t+K {\textsf X~}-\Omega t+k_{2} x_{2}-\omega_{2}t)], \notag
\label{eq:photonstate}
\end{eqnarray}

\begin{eqnarray}
\Psi_{{\bf 2}}\propto \exp[i (k_{{\bf 1r}} x_{1}-\omega_{{\bf 1r}}t+K_{\bf 1r} {\textsf X~}-\Omega_{\bf 1r}t+k_{2} x_{2}-\omega_{2}t)], \notag
\label{eq:photonstate1}
\end{eqnarray}
where the photon and mirror energies and momenta are given by $\hbar \omega$, $\hbar k$, and $\hbar \Omega=M c^{2}/\sqrt{1-(V/c)^{2}}$, $\hbar K=M V/\sqrt{1-(V/c)^{2}}$, respectively. The results from correlated closed-closed photon-atom interferometers presented in sec. \ref{sec:literature} match well this description of the atom-photon system.

Expressions for photon-mirror-photon energy eigenstates {\bf 3}, {\bf 4}, and {\bf 5} follow in a similar manner. The reflected wavevectors and frequencies of these states are calculated using conservation of momentum and energy for elastic reflection. For example, the needed parameters for state {\bf 2} are obtained via eqns. $\hbar k_{1}+\hbar K=-\hbar k_{\bf 1r}+\hbar K_{\bf 1r}$ and  $\hbar \omega_{1}+\hbar \Omega=\hbar \omega_{\bf 1r}+\hbar \Omega_{\bf 1r}$, where $\omega_{1}=k_{1} c$ and $\omega_{\bf 1r}=k_{\bf 1r} c$ are the left side incident and reflected photon frequencies before the right side photon reflects.

The boundary condition, that the three-body wavefunction vanish at the mirror, is then satisfied in a manner similar to that for the reflection of non-zero rest mass particles. The result to lowest order in $V/c$ is
\begin{multline}
PDF_{{\bf tot}}^{eigenstate} \propto \frac{3}{2}-\cos[\alpha]
+\frac{\cos[\alpha-\beta]}{2}
-\cos[\beta],
\label{eq:photonPDF}
\end{multline}
where
\begin{eqnarray}
\alpha=2k_{1}(x_{1}-{\textsf X})~\mbox{and}~\beta=2k_{2}(x_{2}-{\textsf X}).\notag
\label{eq:photonalphabeta}
\end{eqnarray}
  
Note the similarity to that found for massive particles described by eqn. \ref{eq:EntangledInterference}. Eqn. \ref{eq:photonPDF} is independent of the mirror mass while the fringe spacing is determined only by the photon (not the mirror) momenta for the same reasons described in section \ref{sec:reflection} for reflection of massive particles .

Experimental confirmation of these three-body effects is more feasible using photons, particularly due to the availability of low loss mirrors and the ability to produce single photon states. The coherence properties of such states have also been studied \cite{wang,jelezko}. The fixed location of a photon near the mirror, as described in verifying the one-body marginal PDF of eqn. \ref{eq:marginalx1}, can also be determined by measuring the position of the other photon of a pair generated from parametric downconversion. Since this photon pair is correlated before interacting with the mirror, this  is an example of a type C quantum correlation interferometer.

A simple modification of this three-body system is to replace the mirror with a beamsplitter. For non-zero rest mass particles, this is modeled by a delta function potential in the Schr\"odinger equation, $\beta \delta[x_{1}-x_{2}]$, which has its reflectivity related to $\beta$. The boundary conditions, continuity across the boundary but not the first derivative, are satisfied by constraining the the coefficients of the wavefunction resulting in an analytic solution.  Practical realization will more likely involve photons for which low loss beamsplitters are readily available. A beamsplitter allows for more complicated reflection geometries while introducing the ability to vary the strength of the correlation depending on the  beamsplitter reflectivity. 

A potentially more feasible application is with reflection of coherent photon states. However, this introduces an issue associated with the photon number then not being constant. The number of ways that these photons can reflect is ill-defined for a coherent state. A related issue is whether the identical photons in a coherent state reflect as one particle  yielding a simplified calculation of correlation between these photons and the mirror.  A more detailed treatment is beyond the scope of this work.

\section{Transforming open to closed interferometers}
\label{sec:coupled}

Converting an open to a closed substate interferometer dramatically reduces the difficulty in performing correlation experiments. To illustrate such modifications examples are discussed in which attempts are made to generate a closed interferometer for one or all of the substate interferometers which were previously open. To simplify the calculations some interferometer components are treated as classical potentials.

\subsubsection{Two-body interferometers: reflection from opposite sides of the mirror}
\label{sec:oppositesides}

\begin{center}
\begin{figure}
\includegraphics[scale=0.3]{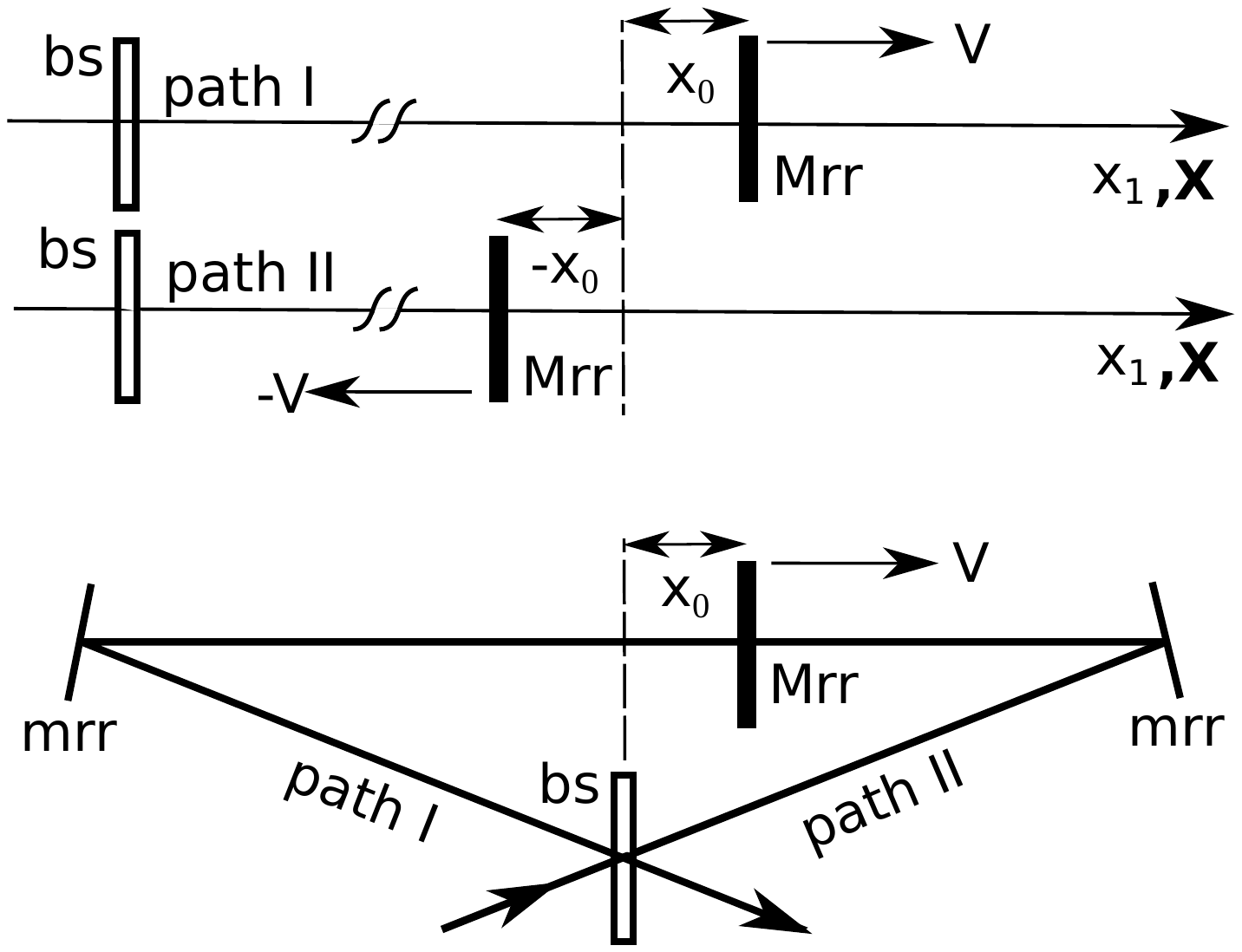}
\caption{Two body reflection from opposite sides of the mirror. The interferometer is shown in the lower half while the upper half indicates how paths I and II can be modeled in one dimension as two-body reflection with a mirror offset. Mirrors mrr and beamsplitter bs are classical potentials while the mirror Mrr is a quantum object. The origin of the coordinates for both the particle and Mrr is at the dashed line.}
\label{fig:2bodyoppositeside}
\end{figure}
\end{center}

The first example attempts to construct a closed interferometer for the particle in the two-body reflection shown in fig. \ref{fig:2body}. To do so, consider the system shown schematically in the lower half of fig. \ref{fig:2bodyoppositeside}. The particle initially traverses a static beamsplitter (a classical potential). The split paths are then directed, via static mirrors (classical potentials), to retro-reflect from opposite sides of a moving mirror, which is treated as a quantum object. The return particle paths then intersect again at the same beamsplitter, or interferometer output port, while the trajectory of the quantum mirror is uninterrupted by classical potentials. 

The particle states beyond the output port, which reflected from opposite sides of the mirror, now move in the same direction. The two-body system shown in fig. \ref{fig:2body} can be modified, as shown in the upper half of fig. \ref{fig:2bodyoppositeside}, to account for the resulting interference. Let the origin of the coordinate systems describing the motion of both the particle and quantum mirror be at the symmetry (or equal arm) position of the particle interferometer. Also, let the quantum mirror be initially offset from this origin by a distance $x_{0}$ while moving at speed $V$. The model of these two paths involves the particle reflecting from the same side of the mirror but with the mirror moving in opposite directions and with opposite initial offsets. There is no initial offset of the particle substates along both paths since they start in the same position, at the beamsplitter (which would be located to the left of the mirror in fig. \ref{fig:2body}). These particle substates then end to the left of the beamsplitter shown in the upper half of fig. \ref{fig:2bodyoppositeside} moving to the left.

Assume that the particle has a non-local spatial extent much greater than $4x_{0}$ while the coherence length of the mirror is larger than the center of mass separation of the two mirror substates which have reflected the particle from opposite sides of the mirror. As discussed in sec. \ref{sec:marginal}, an approximation for the PDF is then given by a superposition of non-local particle and mirror two-body states.

The calculations of the many-body states described in sec.
\ref{sec:theory} involve incident and reflected wavegroups which collide at the origin when $t=0$. To include the mirror offset $x_{0}$ shown in the upper half of fig.  \ref{fig:2bodyoppositeside} this calculation is modified as described in appendix \ref{sec:appendixa}. The resulting two-body PDF which describes measurements of the mirror and the particle beyond the output port of its interferometer is then
\begin{eqnarray}
PDF^{eigenstate} \propto \cos^{2} [\frac{4 m_{1} M (v_{1} {\textsf X}+V x_{1})}{\hbar (m_{1}+M)}].
\label{eq:ringphase}
\end{eqnarray}

The offset $x_{0}$ does not appear in this result. A heuristic explanation involves the initial mirror phase term, $K x_{0}=M V x_{0}/\hbar$, which appears to be different for the two paths due to the positive mirror offset in one and negative in the other. However, the mirror moves in a negative direction in the state with a negative offset, resulting in a negative wavevector and therefore this phase shift term is the same for both states. It then cancels in the cross term of the PDF. 

The superposed particle substates have different momenta as do the superposed mirror substates. Therefore, this is still an open-open interferometer, which is manifest in a phase difference that depends on the particle and mirror coordinates, as shown in eqn. \ref{eq:ringphase}.

For a static mirror the particle states have the same momentum beyond the interferometer output port while the mirror is given momentum kicks in opposite directions yielding a closed-open interferometer. The PDF then depends on the mirror coordinate, which is indicative of the open nature of the mirror substate interferometer. The lack of PDF dependence on when or where the particle is measured indicates a closed particle substate interferometer. For a static mirror localized at position ${\textsf X}_{0}$ interference in a marginal PDF for only a measurement of the particle does not wash out. It results in particle interference with the expected phase $4 m v {\textsf X}_{0}/\hbar$.

For a non-local static mirror the marginal PDF associated with measuring only the particle yields no interference since this involves an integration of eqn. \ref{eq:ringphase} over ${\textsf X}$. However, correlated interference remains. For a microscopic particle reflecting from a mesoscopic static mirror, a measurement of both the mirror at a position which satisfies $n \pi=4m v {\textsf X}/\hbar$ for integer $n$ and a measurement of the particle, at any position beyond the output port, will never occur due to correlated destructive interference in this two-body PDF. 

Consider asynchronous measurement of the non-local particle first, beyond the output port, then the mirror later in this open-closed interferometer. After measurement of the particle but before measurement of the mirror, the mirror is in a one-body superposition state of having reflected the particle from both of its sides. This occurs even for a localized mirror state. The one-body mirror wavefunction, generated after the particle has been measured, is then constructed from the two-body wavefunction by fixing the time and position coordinates of the particle where and when it was measured. 

The experimental advantage in using such an open-closed interferometer in asynchronous measurement is that the particle can be measured at any position beyond the output port while maintaining the one-body mirror superposition state since the particle interference is not spatially localized, as it is for the open-open interferometer shown in fig. \ref{fig:2body}. Nevertheless, the phases of the resulting one-body mirror states are affected by the location of the particle measurement since the time and position coordinates of this particle are thereafter fixed in the two-body wavefunction which is used to construct the one-body mirror state after measurement of the particle.

\subsubsection{Three-body interferometers: reflection from opposite sides of the mirror}
\label{sec:3bodyoppositesides}

Consider now using this interferometer, in which the particle reflects on opposite sides of the moving mirror, to convert the {\em mirror} substate open interferometer into one which is closed. One method to accomplish this involves injecting a delayed second particle, of the same mass and speed as the first, along the same path as the first. The superposition is then of two three-body states. There are again multiple ways such a three-body reflection can occur. The two of interest are: (1) the first particle reflects from one side of the mirror, slowing it down, while the second reflects later from the ``other" side, speeding it up, and (2) the first particle reflects from the ``other" side of the mirror, in which case the mirrors speed increases, while the second particle reflection slows it, again by essentially the same amount for microscopic particles reflecting from a mesoscopic mirror. After reflection of both particles the mirror is in a superposition of states which have the same momentum, yielding a closed interferometer substate. 

There are also other paths in this three-body interferometer, such as ones in which each particle sequentially reflects from the same side of the mirror, which can generate terms in the PDF corresponding to an open mirror interferometer. Under the appropriate conditions, these can vanish in a marginal PDF where the mirror is not measured, leaving only PDF terms associated with this closed mirror substate interference. 

\subsubsection{Two-body interferometers: reflection from the same side of the mirror}
\label{sec:2sameside}

\begin{center}
\begin{figure}
\includegraphics[scale=0.3]{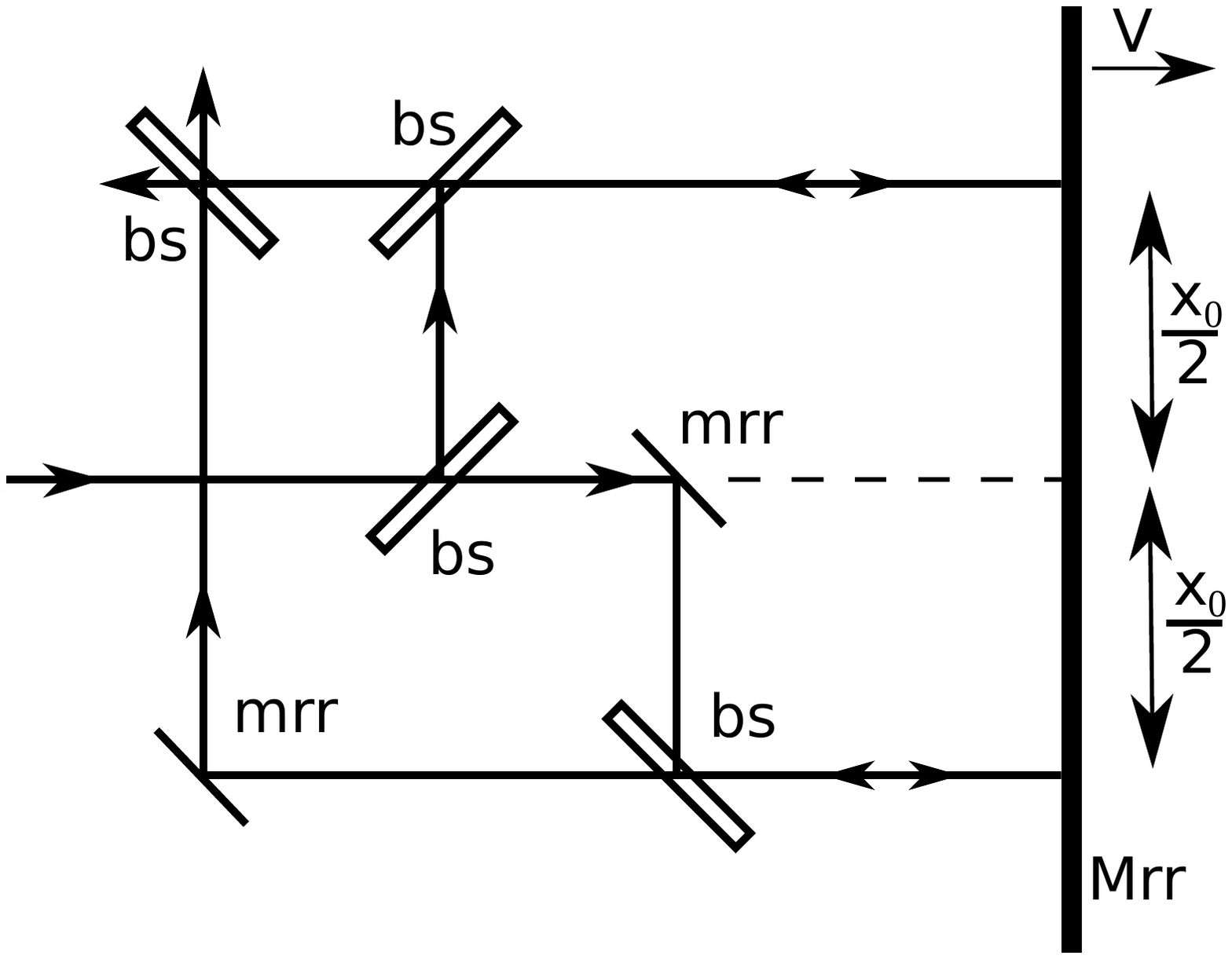}
\caption{Two body reflection from the same side of the mirror. Mirrors mrr and beamsplitter bs are classical potentials while the mirror, Mrr, is a quantum object. The distances along the two paths from the input port bs to Mrr are the same while their return path difference is $x_{0}$.}
\label{fig:2bodysameside}
\end{figure}
\end{center}

The next example involves converting the open-open two-body system shown in fig. \ref{fig:2body} to a closed-closed interferometer. The complexity of the interferometer shown in fig. \ref{fig:2bodysameside}, chosen for this purpose, is offset by the simplicity in calculating the two-body states beyond its output ports. 

The distances along both paths from the input port of the particle interferometer beamsplitter to the quantum mirror are the same.  The two retro-reflected particle states then travel straight through each redirecting beamsplitter (classical potentials).  A mirror (a classical potential) then redirects the lower horizontal path to intersect the output port beamsplitter (a classical potential) while the upper retro-reflected horizontal path goes directly to this beamsplitter. All paths are either horizontal or vertical with perpendicular deflections. Therefore the difference in distance along these two return paths from the mirror, $x_{0}$, is the vertical separation between the two beams where they reflect from the quantum mirror.

The salient feature of this interferometer is that the incident particle wavegroup substate reflects from the quantum mirror at the same time and in the same direction along both paths. Therefore the displacement of the mirror substates associated with recoil along these two paths are the same, resulting in no constraint on the coherence length of the mirror to generate overlap of the mirror substates. The particle substates are assumed to have a non-local spatial extent much larger than $x_{0}$.

The conditions for two-body correlated interference, superposition and overlap, are then satisfied. Correlated interference beyond the output ports can then be approximated by a superposition of two non-local particle and mirror states. One is the reflected two-body state $\Psi[x_{1},{\textsf X}]$ while the other is $\Psi[x_{1}-x_{0},{\textsf X}]$. This superposition is of states each with the same mirror momentum and the same particle momentum, yielding a closed-closed interferometer. For a microscopic particle reflecting from a mesoscopic mirror
\begin{eqnarray}
PDF \approx   \cos ^{2}[\frac{m_{1}(v_{1}-2V)x_{0}}{\hbar}].
\label{eq:samesidereflection}
\end{eqnarray}
The mirror mass enters this expression for $m_{1} \approx M$. It is assumed that $v_{1}>2V$, which is required for retro-reflection from the moving mirror.

This result does not constrain interference for any marginal PDF since there is no dependence on either the particle or mirror coordinates. For example, an indirect measurement of only the particle will show interference even if the unmeasured mirror is in a non-local state. Also, correlated interference does not depend on the positions of the particle or mirror but only on $x_{0}$. That is, if the particle destructively interferes for a given value of $x_{0}$ then so does the mirror, independent of where measurements of the particle (beyond its interferometer outport port) and the mirror are made.

\subsubsection{Three-body interferometers: reflection from the same side of the mirror}
\label{sec:3sameside}

Finally, consider two such fig. \ref{fig:2bodysameside} interferometers slightly displaced vertically from each other, one for each of two distinct particles that retro-reflect from the same side of a quantum mirror which is the only common component of these interferometers. The particle $1$ interferometer has one path delayed, with respect to the other, by a distance of $x_{01}$ {\em only after reflection} from the quantum mirror while that for particle $2$ is $x_{02}$. 

The same assumptions are made as in sec. \ref{sec:2sameside}. In addition, these microscopic particles do not interact with each other. Each particle has it's respective non-local spatial extent much greater than $x_{01}$ and $x_{02}$. This results in the needed overlap of the particle substates for correlated interference as well as the overlap of the mirror substates as discussed in appendix \ref{sec:appendixb}.

Consider the superposition of two three-body states beyond the output ports of the interferometers. Let three-body state A correspond to the paths where particle $2$ reflects from the quantum mirror before particle $1$. However, particle $2$, on return to its output beamsplitter traverses the path with no delay while particle $1$, on return to its output beamsplitter, travels the path with delay $x_{01}$. Three-body state B corresponds to the paths where particle $1$ reflects from the quantum mirror before particle $2$. However, particle $1$ on return to its output beamsplitter now traverses the path with no delay while particle $2$ on return to its output beamsplitter now travels the path with delay $x_{02}$. 

Rather than list all the possible permutations associated with the different reflection sequences along the different interferometer paths, consider the cross term in the PDF associated with interference of states A and B, $\cos[\phi_{A}-\phi_{B}]$. State A is $\Psi_{{\bf 5}}$ modified by the substitution $x_{1} \rightarrow x_{1}- x_{01}$ and ${\textsf X} \rightarrow {\textsf X}_{0}$. State B is $\Psi_{{\bf 4}}$ modified by the substitution $x_{2} \rightarrow x_{2}- x_{02}$ and ${\textsf X} \rightarrow {\textsf X}_{0}$. Each of the three bodies then superposes in the following two ways: particle $1$ travels both with and without delay $x_{01}$, particle $2$ travels both with and without delay $x_{02}$, and the mirror has reflected the particles along paths with and without these delays to their respective output ports. The corresponding cross term in the PDF is
\begin{eqnarray}
PDF_{AB} \approx  \cos ^{2}[\frac{ m_{1}(v_{1}-2V)x_{01}}{\hbar}-\frac{ m_{2}(v_{2}-2V)x_{02}}{\hbar}].
\label{eq:3bodysamesidereflection}
\end{eqnarray}
However, reflection of particles with masses comparable to that of the mirror introduces terms in this PDF which are associated with the positions, masses, and speeds of all three bodies, thereby affecting marginal PDFs.

In spite of eqn. \ref{eq:3bodysamesidereflection} having no such coordinate dependence, correlated interference not only remains but is more easily verified than it would be in an open interferometer. For example, measurement of only particle $1$ anywhere beyond its interferometer output port yields fringes which depend on the variation of parameters associated with the particle $2$ interferometer, such as its path difference $x_{02}$, but which particle $1$ did not traverse. This indirect measurement of the many-body superposition state involves no measurement of either particle $2$ or the mesoscopic mirror. It is manifest in a correlation between the path difference $x_{02}$ (which is a distance between classical objects) and a measurement of particle $1$ at the output port of its interferometer. Yet it still requires quantum behavior of the mesoscopic mirror to mediate this correlated interference via the non-local behavior of the particles. 

Similarly, variation of $v_{1}$ results in interferometric correlations between measurements of particle $2$ anywhere beyond its interferometer output port and the classical device which varies the speed of $m_{1}$ before it is injected into the interferometer. Again, no direct measurement of either particle $1$ or the mesoscopic mirror is necessary. 

\subsubsection{Towards more practical measurements}
\label{sec:practical}

\begin{center}
\begin{figure}
\includegraphics[scale=0.3]{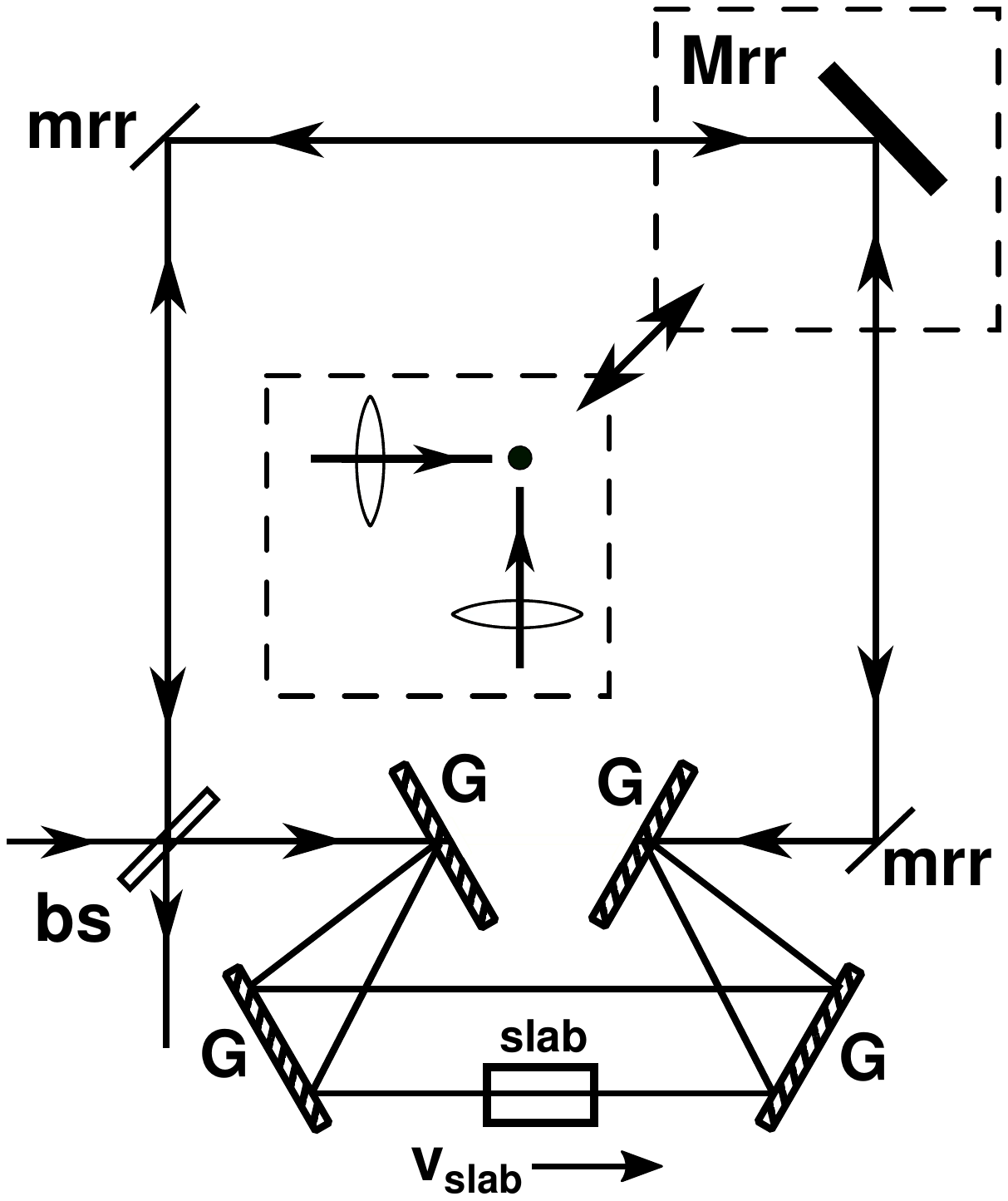}
\caption{Sagnac interferometer schematic for reflection of collinear photons of different wavelengths from the quantum mirror Mrr while all other components are classical potentials. Reflection from atoms or mesoscopic particles is facilitated via replacement of the Mrr with the inset shown in the interior of this Sagnac interferometer. The atoms or mesoscopic particles, indicated by the solid circle inside this inset, move perpendicular to the figure.}
\label{fig:8}
\end{figure}
\end{center}

It is not within the scope of this paper to propose a practical experimental design to measure correlated interference in reflection. The interferometer schematics shown above are intended only as simple illustrations of the fundamental principles. 

Nevertheless, as an example of a more practical system consider three-body reflection from the same side of the mirror using the interferometer of fig. \ref{fig:2bodysameside} with the Mrr initially at rest. Instead of reflecting massive particles, two photons of different wavelengths and different orthogonal linear polarizations traveling collinearly along the paths of this interferometer are used. By placing a waveplate in the return beam path between the lower bs and mrr, the path associated with the photon of one polarization can be varied independently of the other. The three-body PDF of eqn. \ref{eq:3bodysamesidereflection} then becomes $PDF_{AB} \propto \cos^{2} [k_{1} (x_{0}+\delta)-k_{2} x_{0}]$ where $k_{1}$ and $k_{2}$ are the reflected wavevectors of the photons for the two polarizations and $\delta$ is the additional optical path introduced only for the $k_{1}$ photon due to the waveplate. The two photons can be separated beyond the output port either spectrally or via polarization. For a quantum mirror, variation of $\delta$ is then correlated with an interferometric variation of the PDF at the output port for the $k_{2}$ photon even though its path through the interferometer is constant. A measurement need not be made of the Mrr or of the photon with wavevector $k_{1}$. 

One advantage of this system is that reflection from a plane mirror yields a large scattering amplitude compared with that of reflection from an atom or a spherical particle of mesoscopic mass. However, the alignment of a mesoscopic mirror to generate interference is then non-trivial.

A method better designed to facilitate correlated interference in scattering either from atoms or from approximately spherical mesoscopic bodies involves the Sagnac interferometer shown in fig. \ref{fig:8} which is traversed by photons of wavevectors $k_{1}$ and $k_{2}$, that again travel collinearly. To better understand how this interferometer functions consider first reflection from the quantum mirror Mrr .

This Sagnac interferometer is comprised of classical components which are a beamsplitter (bs) two mirrors (mrr) four gratings (G) which spatially separate the photons within the interferometer, and a glass slab of optical path length $nL$ where $n$ is the refractive index and $L$ the slab length. Elastic reflection of the counter-propagating beams from the same side of the Mrr imparts equal momenta to the Mrr as is the case in fig. \ref{fig:2bodysameside}. Only the $k_{1}$ counter-propagating photon states traverse the slab, which moves at speed $v_{slab}$. The counter-propagating states then experience different path lengths due to Fresnel drag which yields a phase difference between them of $\delta \approx k_{1} v_{slab}(n-1)L/c$ \cite{kowalski}. The three-body PDF given in eqn. \ref{eq:3bodysamesidereflection} then becomes $PDF_{AB} \propto \cos^{2} [k_{1} \delta]$.

The ``bias" due to Fresnel drag can also be generated in ways which are less pedagogically appealing. However, most yield small phase shifts. A method to dramatically increase this bias is to replace the slab in fig. \ref{fig:8}  with an optical fiber using the apparatus which is described in ref. \cite{kowalski}. A similar but static slab or optical fiber, along with the two grating pairs, can be inserted in the upper interferometer segment to generate equal paths from the bs to the Mrr for both the $k_{1}$ and $k_{2}$ photons. 

These two photons can be separated beyond the output port spectrally. For a quantum mirror, variation of $\delta$ is then correlated with variation of the PDF for the $k_{2}$ photon even though its path length through the interferometer does not vary. A measurement then need not be made of the Mrr or of the photon with wavevector $k_{1}$ to observe this correlation.

To modify this system to measure correlated interference in reflection for atoms (or mesoscopic scatters) consider replacing the dashed box around the Mrr in fig. \ref{fig:8} with the dashed box that is shown inside the Sagnac interferometer. The two lenses focus the photon beams onto the solid circle which represents an atom moving in a direction perpendicular to fig. \ref{fig:8}. The Rayleigh scattered photons must be emitted into a small solid angle to generate interference, resulting in a reduced scattering amplitude compared with that of the Mrr. The fig. \ref{fig:2bodysameside} and \ref{fig:8} interferometers are far from a comprehensive treatment of the diversity of experimental designs possible to verify quantum correlation interferometry in reflection.

\section{Summary}
\label{sec:summary}

The above calculations of three-body quantum correlation interferometry demonstrate that measurement of either one or both of the microscopic probes can reveal the quantum behavior of the mirror without a direct measurement of it. As this mirror increases from a microscopic to a mesoscopic mass the quantum-classical transition occurs when such correlated interference disappears. The mirror then behaves as a classical object which cannot be in a superposition of states which correspond to different reflection sequences. Such interferometric correlation does not exist without this superposition. 

These results are perhaps better understood by first considering the similarities and differences between the two-body correlation interferometer in reflection, shown in fig. \ref{fig:2bodysameside}, and the correlated photon-atom interferometer described by Tomkovic et. al. \cite{tomkovic} which was outlined in section \ref{sec:literature}. In particular, consider the mirror in the reflection system to be analogous to the atom in the photon-atom system. Both systems are closed-closed interferometers, both incorporate classical potentials to deflect and combine the beams, and both use conservation of energy and momentum in recoil to generate the correlation. Both yield interference in one substate interferometer when only the path difference of the other substate interferometer is varied. One involves a photon emitted from an atom while the other utilizes a photon reflected from the atoms in a mirror. An important difference is the ease with which the atom can be measured as opposed to the center of mass of the mirror, particularly as the mirror mass is varied from micro to mesoscopic. 

Reflecting two microscopic particles from the mirror in the three-body correlation interferometers, described in sections \ref{sec:3sameside} and \ref{sec:practical}, eliminates the need to measure the mirror while maintaining a similar interferometric correlation between the two particles. That is, a variation in the path length of one particles interferometer generates interference in the other particle even though the path difference in its interferometer does not change. This correlated interference is mediated by the mirror which is in a superposition state associated with the order of particle reflection being indeterminate.

The difficulty of experimentally measuring the quantum-classical transition is mitigated in using such a three-body system for the following reasons: (1) the mesoscopic quantum mirror traverses neither a division of amplitude (the mirror does not have to pass through a beamsplitter) nor a division of wavefront interferometer (the mirror does not have to fit through slits), (2) there is little constraint on alignment of the interferometer to generate the superposed mirror states, (3) the interferometric constraints on the mirror's coherence length are reduced due to the small (or non-existent, as discussed in sec. \ref{sec:2sameside}) center of mass displacement between these superposed mirror states, (4) this small displacement reduces environmental decoherence of the mirror's superposition state, (5) indirect measurement of the mirror's quantum behavior is possible in a closed-closed-closed interferometer, (6) while much of the discussion above dealt with non-zero rest mass particle reflection, photon-mirror correlation, which is fundamentally related to photon-atom correlation, is predicted to exhibit similar interferometric effects. Low loss mirrors and beamsplitters for such an application are ubiquitous. (7) The path difference in one particle substate interferometer needed to cycle through one fringe in the other particle substate interferometer, as described in sec. \ref{sec:3sameside}, is roughly $2 \pi \hbar/m_{1} v_{1}$, which can be a mesoscopic distance. This interference reveals the quantum behavior of the mirror of mass $M$. However, if a body with the mirror's mass were to traverse a Michelson interferometer comprised of classical potentials then an interferometer path difference of $2 \pi \hbar/M V$ would yield one fringe at its output port. Such a small path difference, needed to verify the quantum behavior of the mesoscopic mirror with this interferometer, would be difficult to measure due to the required dimensional stability of the interferometer. Although insensitive to such changes in path difference, correlation interferometry described above for microscopic particles reflecting from mesoscopic mirrors nevertheless maintains interference. (8) Interferometric correlation between a measurement of only one microscopic particle at its interferometer output port and the path difference of the other particle's interferometer reveals the superposition state of the mesoscopic mirror as described in sec. \ref{sec:3sameside}. There is neither a need for a correlation measurement between the two particles nor between them and the mirror.

The foundations upon which this analysis of the quantum-classical boundary is based are quantum correlation in particle-mirror recoil and environmental decoherence which depends on the small center of mass separation of the superposed states. On a microscopic scale both of these fundamental issues have been experimentally confirmed. 

Such a narrow focus on the quantum-classical transition and on only two and three-body systems provides limited insight into the breadth of quantum correlation interferometry. However, possible applications of both type B and C systems rely on establishing the quantum-classical boundary to determine the regime in which this correlation is maintained.

\appendix 

\section{Two-body delayed reflection}
\label{sec:appendixa}

The peaks of the incident and reflected Gaussian wavegroups, as presented in sec. \ref{sec:theory}, meet at the origin when $t=0$. Consider now the necessary modifications to these results for the peaks to overlap at offset positions and times. The first constraint is for the motion of these peaks to correspond with those of classical objects and that they reflect as such to satisfy the expected behavior when the states are localized. That is, the trajectories of the incident and reflected offset two-body wavegroups must overlap at the appropriate classical collision position and time in the $(x_{1},{\textsf X~},x_{2})$ plane. The second constraint, needed for non-local states, is that the resulting two-body wavefunctions must satisfy the boundary condition.

Rather than treating such offsets in a three-body system consider the two-body correlated interferometer in reflection shown in fig. \ref{fig:2body}. The ``classical trajectory" and wave boundary constraints are satisfied with the following substitutions which are made into the incident two-body wavefunction that is given by eliminating particle $2$ in eqn. \ref{eq:state1}: ${\textsf X~}\rightarrow {\textsf X~}+{\textsf X~}_{i0}$ and $x_{1} \rightarrow x_{1}+x_{i0}$, where ${\textsf X~}_{i0}$ and $x_{i0}$ correspond to the offsets of the mirror and particle $1$ at $t=0$. In addition, the following substitutions are made into the reflected two-body wavefunction which is given by eliminating particle $2$ in eqn. \ref{eq:state2}: ${\textsf X~}\rightarrow {\textsf X~}+{\textsf X~}_{r0}$ and $x_{1} \rightarrow x_{1}+x_{r0}$.

Two applications are discussed next. First, the mirror is offset initially by position $x_{0}$ with no particle offset. In this case ${\textsf X~}_{i0}=x_{0}$ and $x_{i0}=0$ while ${\textsf X~}_{r0}=(M-m_{1})x_{0}/(M+m_{1})$ and $x_{r0}=2 M x_{0}/(M+m_{1})$. This offset is applied to the interferometer described in sec. \ref{sec:oppositesides}. Second, the particle is offset initially by position $x_{0}$ with no mirror offset. In this case ${\textsf X~}_{i0}=0$ and $x_{i0}=x_{0}$ while ${\textsf X~}_{r0}=2 m_{1} x_{0}/(M+m_{1})$ and $x_{r0}= (m_{1}-M)x_{0}/(M+m_{1})$. These substitutions, applied to expressions for two-body reflection of Gaussian wavepackets, result in the expected offset classical trajectories of the wavegroup substate peaks while also satisfying the boundary condition.

\section{Three-body delayed reflection from the same side of the mirror}
\label{sec:appendixb}

Consider now the effect of the particle $1$ wavegroup substate peak reflecting from the mirror later than that of particle $2$ on the difference in mirror wavegroup peak positions for interferometer states A and B, discussed in sec. \ref{sec:3sameside}. States A and B correspond to either particle $2$ reflecting from the mirror before particle $1$ or this reflection order reversed. As discussed in sec. \ref{sec:wavegroups} the order of reflection of the wavegroup peaks is not related to the reflection order of the particles when the wavegroups overlap. If the A and B mirror substate wavegroup peaks are offset by a distance much greater than their coherence lengths, due to a delay in the reflection of particle substate peaks $1$ and $2$, then overlap of the mirror substate wavegroups and therefore interference is eliminated.

However, there is no difference in the positions of the particles and mirror wavegroup substate peaks between states A and B during reflection. The difference between states A and B occurs only after reflection on the return paths to the particle output port beamsplitters. Therefore such a delay in the sequence at which the two particle wavegroup peaks reflect does not generate a difference in mirror displacement between states A and B.


\begin{references}
\bibitem{classical} The term classical, in this context, is used for beamsplitters and mirrors of an interferometer which simply split or reflect the wavefunction. There is no uncertainty in the positions nor any correlation associated with recoil effects in the interactions of these components with the wave traversing the interferometer.
\bibitem{beams} Of course two plane waves intersecting at an  angle also have a direction along which the PDF is constant. However, for beams, whose transverse PDF profile is limited, there is no direction.
\bibitem{mixture} The other possibility is that the mirror collapses into a random or statistical mixture of positions commensurate with the two-body Gaussian distribution at the time that the particle is measured. This is inconsistent with measurements, which have insufficient resolution to distinguish between states, resulting in a superposition rather than a mixture.
\bibitem{kowalski3} More realistic models of reflection can be used but they add complexity and limit generality.
\bibitem{comment} Quantum correlation in a simultaneous three-body collision is then determined by more than just the conservation laws.
\bibitem{beamsplitter} For non-zero rest mass particles, this is modeled by a delta function potential in the Schr\"odinger equation, $\beta \delta[x_{1}-x_{2}]$, which has its reflectivity related to $\beta$. The boundary conditions, continuity across the boundary but not the first derivative, are satisfied by constraining the the coefficients of the wavefunction, resulting in an analytic solution.
\end{references}
\end{document}